\documentclass[journal=mamobx,manuscript=article]{achemso}

\usepackage[version=3]{mhchem} 
\usepackage{amssymb}
\usepackage{xcolor}
\usepackage[colorlinks]{hyperref}

\usepackage{multirow}
\usepackage{booktabs}

\renewcommand{\figurename}{\textbf{Figure}}
\makeatletter
\renewcommand{\fnum@figure}[1]{\textbf{\figurename~\thefigure.}}  
\makeatother

\author{Xianggui Zhou}
\affiliation[South China University of Technology]
{South China Advanced Institute for Soft Matter Science and Technology, School of Emergent Soft Matter, South China University of Technology, Guangzhou 510640, China}
\author{Nengjie Cao}
\affiliation[South China University of Technology]
{South China Advanced Institute for Soft Matter Science and Technology, School of Emergent Soft Matter, South China University of Technology, Guangzhou 510640, China}
\author{Xiang-Meng Jia}
\affiliation[South China University of Technology]
{South China Advanced Institute for Soft Matter Science and Technology, School of Emergent Soft Matter, South China University of Technology, Guangzhou 510640, China}
\author{Jinyuan Mao}
\affiliation[South China University of Technology]
{South China Advanced Institute for Soft Matter Science and Technology, School of Emergent Soft Matter, South China University of Technology, Guangzhou 510640, China}
\author{Jiajia Zhou}
\email{zhouj2@scut.edu.cn}
\affiliation[South China University of Technology]
{South China Advanced Institute for Soft Matter Science and Technology, School of Emergent Soft Matter, South China University of Technology, Guangzhou 510640, China}
\alsoaffiliation[]
{Guangdong Provincial Key Laboratory of Functional and Intelligent Hybrid Materials and Devices, South China University of Technology, Guangzhou 510640, China}
\alsoaffiliation[]
{State Key Laboratory of Pulp and Paper Engineering, South China University of Technology, Guangzhou 510640, China}

\title[]{Size Effect of Monovalent Ions on Polyelectrolyte Brushes}

\begin{document}

\begin{abstract}

The conformation of polyelectrolyte (PE) brushes is highly sensitive to external conditions, particularly salt concentration and ion-specific effects. 
As salt concentration increases, PE brushes transition from an osmotic brush regime at low salt ($H \propto c_\mathrm{s}^{0}$) to a salted brush regime at high salt ($H \propto c_\mathrm{s}^{-1/3}$).
However, deviations from this ideal scaling behavior are frequently observed in molecular simulations. 
In this work, we employ coarse-grained molecular dynamics simulations to systematically investigate how the sizes of counterions and co-ions affect the structural evolution and scaling behavior of PE brushes over a broad range of salt concentrations.
Our results show that counterion size plays a dominant role in regulating ion penetration and coordination with PE monomers. 
At low salt concentration, smaller counterions penetrate more easily into the brush, leading to enhanced local charge compensation and stronger brush collapse. 
At high salt concentration, however, the brush height becomes largely insensitive to counterion size, while deviations from the classical scaling relation emerge.
On the other hand, co-ion size mainly affects the system indirectly by modifying ion distributions and the local electrostatic environment. 
Smaller co-ions weaken local charge compensation and suppress brush collapse, with this effect becoming more pronounced at high salt concentration.
When the sizes of counterions and co-ions are reduced simultaneously, the system exhibits a coupled response. 
At low salt concentration, the behavior remains counterion-dominated. 
At high salt concentration, enhanced ion penetration and ion crowding modify the local ionic environment, leading to reentrant swelling and deviations from classical scaling relations.
Collectively, this work provides a microscopic understanding of how ion size and salt concentration jointly govern the structural response of PE brushes and the emergence of non-classical scaling behavior in realistic solution environments.

\end{abstract}

\section{Introduction}
\label{sec:PEB_Introduction}

Polyelectrolyte (PE) brushes are formed by densely grafting charged polymer chains onto surfaces such as solid substrates, nanoparticles, or the inner walls of nanochannels \cite{das2015polyelectrolyte}. 
Owing to electrostatic repulsion between charged monomers and the osmotic pressure of counterions, the grafted chains extend away from the surface and adopt stretched conformations in solution \cite{ishraaq2024all, pial2022atomistic}.
In PE brush systems, chain conformational entropy, excluded-volume interaction, long-range electrostatic interaction, and the migration and redistribution of mobile ions are strongly coupled.
As a result, the structure and properties of PE brushes are highly responsive to external stimuli, including salt concentration \cite{xu2018structure, mondarte2023unveiling}, ion valency and size \cite{guptha2014polyelectrolyte, xu2019ion}, pH \cite{ionov2004inverse, chen2015electroosmotic}, solvent quality \cite{dimitrov2007polymer, li2025synergistic}, electric field strength \cite{pial2021overscreening, pial2022charge}, and temperature \cite{sachar2021all}. 
Moreover, these responses often exhibit pronounced time dependence and hysteresis effects \cite{hollingsworth2021hysteretic, xiong2023neuromorphic, nekoubin2024improved, nekoubin2024highly}.
These properties endow PE brushes with broad applications, such as surface lubrication enhancement \cite{yu2018multivalent}, emulsion stabilization \cite{saleh2005oil}, antifouling \cite{kobayashi2012wettability, liu2017antifouling} and antibacterial coatings \cite{luo2023antimicrobial}, gene \cite{zhang2012disulfide, chen2019polycationic} and drug delivery \cite{yang2012amphiphilic}, biosensing \cite{ali2008biosensing}, functional nanofluidic ion channels for current rectification \cite{hou2012building, zhou2022dynamically}, and electrochemical devices \cite{schaefer2013high, xie2022spherical}.
Therefore, understanding the structure and conformation of PE brushes is important for both fundamental studies and practical applications. 
Such understanding can guide the rational design and optimization of functional materials.

The effect of salt concentration on the structure of PE brushes has attracted considerable attention for decades.
For neutral brushes, the seminal works of Alexander and de Gennes showed that the brush height $H$ follows a power-law dependence on the degree of polymerization $N$ and the grafting density $\sigma_\mathrm{g}$ ($H \propto N\sigma_\mathrm{g}^{1/3}$) \cite{de1976scaling, alexander1977adsorption}. 
Following this framework, de Gennes, Pincus, and Zhulina \emph{et al.} extended the scaling argument to charged systems, identifying two characteristic regimes \cite{de1976remarks, pincus1991colloid, borisov1991collapse}.
Under low-salt conditions, the system is in the \textit{osmotic brush} regime, where the brush height follows the scaling relation $H \propto N f^{1/2}$, with $f$ denoting the charge fraction of the PE chains. 
In this regime, the brush height is insensitive to the external salt concentration and is dominated by the osmotic pressure of counterions confined within the brush.
Under high-salt conditions, the system crosses over to the \textit{salted brush} regime, where the brush height exhibits a power-law decay with salt concentration $c_\mathrm{s}$, described by $H \propto N f^{2/3}\sigma_\mathrm{g}^{1/3}c_\mathrm{s}^{-1/3}$. 
Chen and co-workers developed a theoretical model for two opposing PE brushes and derived analytical expressions for the scaling relation between brush height and salt concentration under different conditions \cite{duan2023swelling}. 
Li and Yu \emph{et al.} introduced a multivalent-ion adsorption mechanism and proposed a unified theoretical framework to describe the structural response of PE brushes \cite{li2022effects, li2025synergistic}. 
Their model not only reproduces the classical scaling behavior under monovalent salt solutions, including both osmotic and salted brush regimes, but also quantitatively predicts an ion-valency-driven phase transition: when the ion valency exceeds a critical threshold ($z \approx 1.38$), the brush undergoes a pronounced collapse transition.
In addition, Chen and co-workers proposed a new scaling regime based on a cell model, termed the nonoverlapping electric double layer brush (NOEB) \cite{chen2026scaling}:
When the electric double layers surrounding neighboring monomers no longer overlap, the brush height follows a power-law scaling relation with salt concentration $H \propto c_\mathrm{s}^{-1/5}$ at high salt condition.

Relevant experiments have also systematically confirmed the scaling relation $H \propto c_{\mathrm{s}}^\alpha$, where $\alpha$ denotes the scaling exponent.
For example, Tirrell, Yu, and co-workers used surface force apparatus (SFA) and neutron reflectometry (NR) to examine the structural response of strongly charged polystyrene sulfonate (PSS) brushes at different salt concentrations \cite{yu2016structure}.
At low salt concentration, the brush height remains nearly constant and is insensitive to the bulk NaNO$_3$ concentration, consistent with the osmotic brush regime.
In the salted brush regime, the brush height follows a power-law dependence on salt concentration, with crossover concentrations of about 0.03 M in the SFA measurements and 2 M in the NR study. 
The measured exponents are $\alpha=-0.29 \pm 0.04$ from NR and $\alpha=-0.27 \pm 0.01$ from SFA, both close to the theoretical prediction of $\alpha=-1/3$.
Hollingsworth and Larson used quartz crystal microbalance measurements to study the osmotic-to-salted brush transition in sparsely grafted long-chain PAA brushes, revealing that hysteresis depends strongly on chain length and grafting density\cite{hollingsworth2021hysteretic}. 
In their work, the effective scaling exponent $\alpha$ is approximately $+0.13 \sim +0.15$ at low salt concentration and $-0.17 \sim -0.20$ at high salt concentration. 

Molecular dynamics (MD) simulations can provide microscopic details that are often difficult to access experimentally. 
Therefore, MD simulations are important for understanding the molecular mechanism underlying the scaling predictions.
Kumar and Seidel employed Langevin dynamics (LD) simulations combined with the MMM2D method to accurately describe electrostatic interactions, and observed the transition from the osmotic brush to the salted brush regime in monovalent salt solutions \cite{kumar2005polyelectrolyte}. 
At relatively high salt concentrations, the brush height scales with the salt concentration as $H \propto c_\mathrm{s}^{-0.15}$. 
However, when the scaling analysis was based on the total concentration of all mobile ions within the brush, including intrinsic counterions, added salt cations, and co-ions, the scaling exponent $\alpha$ shifted to approximately $-0.31$, close to the classical prediction.
Ibergay \emph{et al.} investigated salt-containing PE brushes with explicit solvent using dissipative particle dynamics (DPD) combined with the particle-particle particle-mesh (PPPM) method for electrostatics \cite{ibergay2010mesoscale}. 
They reported a scaling relation of $H \propto c_\mathrm{s}^{-0.12}$ in the salted brush regime. 
After excluded-volume corrections associated with polymer chains were included, the scaling exponent $\alpha$ was found to recover toward $-0.31$, highlighting the importance of steric interactions in determining the scaling behavior.
Similarly, LD simulations by Guptha and Hsiao yielded a scaling relation of $H \propto c_\mathrm{s}^{-0.15}$ in the salted brush regime \cite{guptha2014polyelectrolyte}. 
This result is consistent with the theoretical prediction of an exponent close to $\alpha=-1/6$ when electrostatic contributions to chain stiffness are neglected and Odijk-type excluded-volume effects are considered \cite{hariharan1998ionic}.
Notably, these coarse-grained (CG) simulation results deviate from the classical scaling law $H \propto c_\mathrm{s}^{-1/3}$. 
Kumar and Ibergay \emph{et al.} argued that this discrepancy primarily arises from finite system sizes and limited range of salt concentrations accessible in simulations, which often prevent reaching the asymptotic high-salt limit \cite{kumar2005polyelectrolyte, ibergay2010mesoscale}.
Thus, the observed scaling exponents, typically around $-0.15$, correspond to a crossover regime between the osmotic and salted brush states. 
In this regime, intrinsic counterions still play an important role, while excluded-volume and chain-elasticity effects limit the collapse induced by electrostatic screening.
In recent work, Miao, Hao, and co-workers employed LD simulations to investigate the coupled effects of solvent quality and multivalent salts on the conformational behavior of PSS brushes \cite{hao2025systematic}. 
Their results show that when the cation size is comparable to the monomer size, the scaling exponent approaches the classical value of $\alpha=-1/3$. 
In contrast, when the ion size is either very small (e.g., diameter $\sim 0.1$) or relatively large (diameter of $> 1.0$), the scaling exponent decreases to $\alpha \approx -0.15$, indicating that ion size introduces pronounced specific effects and modulates the scaling behavior.
These scaling relationships show how the height of a PE brush depends on salt concentration. 
They also provide a key foundation for understanding how the brush structure evolves and how it can be regulated.

Ion-size specificity helps us understand how the microscopic morphology of polyelectrolyte brushes evolves. 
It also serves as a useful tuning parameter for the rational design of stimulus-responsive materials.
Tirrell, Yu, and co-workers demonstrated that even at identical valency, variations in ionic size can significantly alter the spatial organization and structural uniformity of brush layers \cite{farina2015reversible}. 
In particular, \ce{La^{3+}}, because of its small ionic radius and high charge density, exhibits markedly stronger adhesion compared to other trivalent ions. 
For divalent counterions, distinct structural responses are observed: \ce{Mg^{2+}} and \ce{Ca^{2+}} tend to induce a homogeneous brush collapse, whereas \ce{Ba^{2+}} promotes the formation of heterogeneous pinned micelle structures \cite{xu2019ion}. 
Accordingly, the variation in brush height is closely correlated with ionic size.
Kuo \emph{et al.} showed, using quartz crystal microbalance measurements, that the hydrophilicity/hydrophobicity of anions modulates brush dehydration and collapse by tuning hydrophobic interactions within ion pairs \cite{kou2018counterion}. 
The effect follows the order \ce{Ac^-} $<$ \ce{Cl^-} $<$ \ce{SCN^-} $<$ \ce{ClO4^-}. 
This trend indicates that more hydrophobic anions, such as \ce{ClO4^-} and \ce{SCN^-}, more readily induce brush dehydration and collapse, whereas more hydrophilic anions, such as \ce{Ac^-}, promote brush swelling at low salt concentrations. 
This behavior can be attributed to the modulation of local osmotic pressure and interchain interactions during ion-pair formation.
All-atom simulations by the Das group revealed that ion size critically influences bridging interactions \cite{pial2021quantification}. 
In their work, small ions such as \ce{Li+}, have compact solvation shells that cannot simultaneously coordinate oxygen atoms from nonadjacent monomers, resulting in weak bridging. 
In contrast, ions of intermediate size, such as \ce{Na+}, effectively promote both intrachain and interchain bridging. 
Moreover, their subsequent work showed that small ions tend to remain within the brush because of their high charge density, which leads to larger enthalpic gains upon condensation \cite{pial2022specific}. 
Larger ions can also reside within the brush, but they introduce greater entropic penalties because of their larger solvation shells and associated water molecules.
Miao and co-workers used CG molecular dynamics simulations to systematically investigate ion-size effects on brush morphology \cite{tan2023size}. 
They reported a sequence of structural transitions with increasing counterion size, including homogeneous collapse, pinned micelle formation, vertical stratification, and re-stretching, accompanied by a nonmonotonic variation in brush height. 
Their further studies showed that the coupled effects of ion size and salt concentration can induce transitions from uniform collapse to heterogeneous structures and finally to re-swollen states \cite{hao2025systematic}. 
In addition, they proposed that in highly concentrated multivalent salt solutions, the penetration of co-ions into the brush not only maintains local electroneutrality but also promotes brush re-swelling through electrostatic confinement effects.
Faraday \emph{et al.}, using all-atom simulations, demonstrated that smaller co-ions can suppress brush collapse at high salt concentrations by weakening local charge compensation \cite{rodriguez2013ionic}. 
These findings highlight the importance of ion-specific effects beyond mean-field descriptions.

Despite these advances, several unsolved issues remain. 
A systematic understanding of brush height and its scaling behavior with salt concentration for monovalent salt conditions is already established, but the role of ion size in governing these scaling laws remains unclear.
In addition, previous studies have focused mainly on counterion effects, whereas the role of co-ions in regulating brush structure remains poorly understood. This is especially true for how co-ions behave as a function of their size and how they work together with counterions.
Motivated by the above questions, we employ CG molecular dynamics simulations to systematically investigate the effects of ion size on the structure and salt-dependent properties of PE brushes. 

This work is organized into three parts. 
First, the effect of counterion size on PE brush structures and salt-dependent scaling relations is studied. 
Second, the influence of co-ion size on the structural response of PE brushes is investigated. 
Finally, the coupled effect of simultaneously varying counterion and co-ion sizes is explored to clarify their synergistic role in regulating PE brush behavior. 
Overall, we aim to clarify how ion size affects the structural evolution and scaling laws of polyelectrolyte brushes. 
This study provides insight into ion-specific phenomena in functional interfacial materials.

\section{Simulation Model and Methods}
\label{sec:PEB_Method}

\begin{figure}[htbp]
  \centering
  \includegraphics[width=0.8\textwidth]{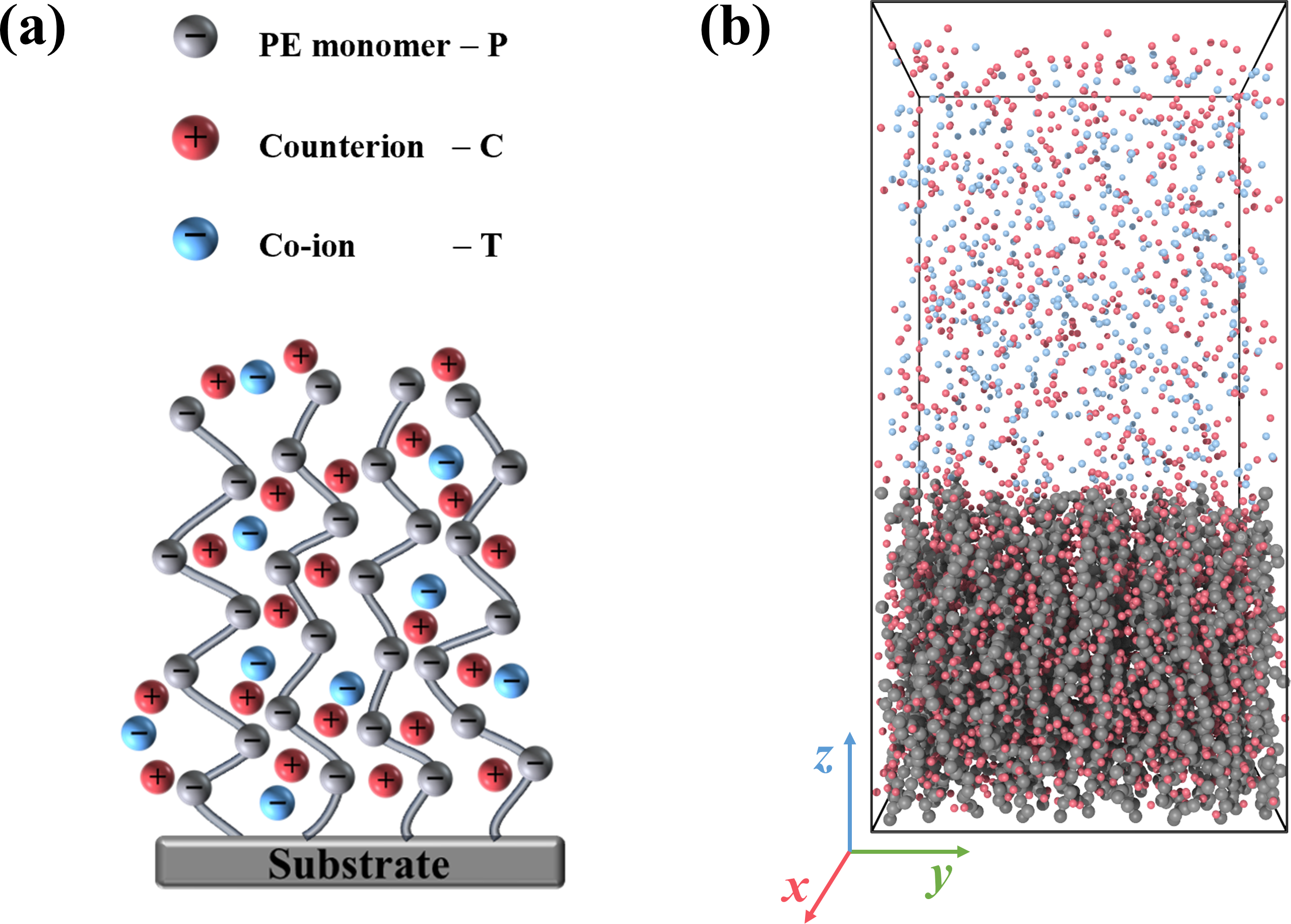}
  \caption{Schematic illustration of the PE brush model. (a) Simplified model, (b) MD simulation snapshot at a salt concentration of $c_\mathrm{s}=0.01$.}
  \label{fig:PEB_Model}
\end{figure}

In this study, we employ a bead-spring model for the PE brush. 
The simulation is performed in a rectangular box of dimensions $L_x \times L_y \times L_z = 30 \times 30 \times 60\,\sigma^3$, where $\sigma$ denotes the reduced length unit.
Periodic boundary conditions were applied in the $x$, $y$, and $z$ directions. 
The grafting density is fixed at $\sigma_{\rm g} = 0.1\,\sigma^{-2}$. 
The system contains $90$ PE chains, and each chain consists of $N = 30$ monomers of diameter $1.0\,\sigma$. 
Following the common interpretation of the Kremer--Grest model, the bead diameter $\sigma$ is taken to be comparable to the Kuhn length of a real polymer chain, which is typically on the order of $\sim1\,\mathrm{nm}$ \cite{rovigatti2019numerical}. 
Therefore, $\sigma$ may be roughly associated with a sub-nanometer to nanometer length scale ($\sim0.5$--$1.0\,\mathrm{nm}$). 
Using PSS as a reference, whose repeat-unit spacing is approximately $0.25\,\mathrm{nm}$ \cite{dobrynin2005theory}, one bead therefore represents several repeat units. 
Accordingly, the simulated chains ($N=30$) may be roughly associated with contour lengths on the order of $15$--$30\,\mathrm{nm}$.
As illustrated in \textbf{Figure~\ref{fig:PEB_Model}}, gray beads represent monomers carrying a unit negative charge $-q$. 
All PE chains are randomly grafted onto the $xy$-plane substrate through a neutral end bead (not shown in the figure and not counted as part of the chain length).
In the simulations, the PE is treated as a strong polyelectrolyte: each monomer fully dissociates and releases one counterion with unit positive charge. 
Typical examples of strong polyelectrolyte brushes include sodium polystyrene sulfonate (NaPSS)\cite{tran1999polyelectrolyte, tran2001synthesis, yu2016structure} and poly(2-acrylamido-2-methylpropane sulfonate sodium) (PAMPSNa) \cite{su2023synthesis}.
In addition, monovalent salt is introduced into the system. 
The salt fully dissociates as cations and anions, which are initially placed randomly throughout the simulation box. 
The salt concentration $c_{\mathrm{s}}$ is defined in terms of number density as $c_\mathrm{s} = N_{\mathrm{salt}} / (L_x L_y L_z)$, 
where $N_{\mathrm{salt}}$ denotes the number of salt ion pairs. 
For clarity, all cations originating from both PE dissociation and salt dissociation are treated as counterions with charge $+q$ and are shown as red spheres in \textbf{Figure~\ref{fig:PEB_Model}}.
The anions from salt dissociation are defined as co-ions with charge $-q$ and are shown as blue spheres.
All system is electrically neutral.

To investigate ion-size effects on PE brush structure, we first constructed a reference system with counterion and co-ion diameters of $1.0\,\sigma$. 
Based on this system, three simulation protocols were considered:
\begin{itemize}
\item[(i)] decreasing the counterion diameter while fixing the co-ion diameter at $1.0\,\sigma$;
\item[(ii)] decreasing the co-ion diameter while fixing the counterion diameter at $1.0\,\sigma$;
\item[(iii)] simultaneously decreasing the diameters of both counterions and co-ions.
\end{itemize}
The scaling behavior predicted in the salted-brush regime is derived under the asymptotic condition that the external salt concentration is much higher than the concentration of counterions released by the PE brush itself \cite{kumar2005polyelectrolyte, ibergay2010mesoscale}. 
Therefore, to examine the scaling behavior, simulations were performed to maximize this concentration ratio. 
Relatively short chains were used to improve computational efficiency and to facilitate access to this asymptotic salted-brush condition.
The maximum accessible salt concentration is constrained by excluded-volume effects, strong electrostatic interactions, and numerical stability, and therefore varies among different ion-size combinations. 
The diameters of ions are varied from $0.3\,\sigma$ to $1.0\,\sigma$. The resulting size range corresponds approximately to $0.15$--$1.0\,\mathrm{nm}$, covering experimentally relevant dimensions from small bare ions to larger hydrated and weakly coordinating ionic species. \cite{nightingale1959phenomenological}
The corresponding simulation protocols, salt concentration ranges, and brush-height data are summarized in \textbf{Tables~S1--S3}. Representative top-view and side-view snapshots of the equilibrated PE brushes are provided in \textbf{Figures~S2, S6, and S9}.

All CG simulations were performed using the GALAMOST package \cite{zhu2013galamost}. The simulations were conducted in the canonical (NVT) ensemble with the implicit solvent. The total potential energy consists of bonded, nonbonded, and electrostatic interactions, as well as the wall potential
\begin{equation}
    U_{\rm total} = U_{\rm nb} + U_{\rm b} + U_{\rm Coul} + U_{\rm wall}
    \label{U_total}
\end{equation}

The nonbonded interactions $U_{\rm nb}$ are described by the standard shifted and truncated Lennard-Jones (LJ) potential, also known as Weeks--Chandler--Andersen (WCA) potential\cite{weeks1971role}:
\begin{equation}
    U_{\rm nb}(r_{ij}) = \left\{ \begin{array}{cl} \displaystyle
         {4\varepsilon \left\lbrack {\left( \frac{\sigma_{ij}}{r_{ij}} \right)^{12} - \left( \frac{\sigma_{ij}}{r_{ij}} \right)^{6}} \right\rbrack + \varepsilon}& , \quad r_{ij}  \leq \sqrt[6]{2}\,\sigma_{ij}\\
        {0}& , \quad r_{ij} > \sqrt[6]{2}\,\sigma_{ij} \end{array} \right.
    \label{eq:LJ}
\end{equation}
in which $r_{ij}$ denotes the distance between two nonbonded particles, and $\varepsilon$ represents the energy scale of the pairwise interaction.
The length parameter $\sigma_{ij}$ is defined as the geometric mean of the diameters of the interacting particles, i.e., $\sigma_{ij} = \sqrt{\sigma_i \sigma_j}$. The cutoff distance is set to $\sqrt[6]{2}\,\sigma_{ij}$.
For nonbonded interactions between non-adjacent beads along the PE chain, $\sigma_{ij}$ is fixed at $1.0\,\sigma$. 
This purely repulsive potential corresponds to good solvent conditions. 

The covalent bonds are described by the finitely extensible nonlinear elastic (FENE) potential:
\begin{equation}
    U_{\rm b}(r) = - \frac{1}{2} k_{\rm b} {r_{\rm max}^{2}} \ln \left( 1 - \frac{r^{2}}{r_{\rm max}^{2}} \right)
    \label{eq:FENE}
\end{equation}
where the spring constant is $k_{\rm b} = 30.0\,\varepsilon/\sigma^2$, and the maximum bond length is $r_{\rm max} = 1.5\,\sigma$.
Under these parameters, the average bond length is approximately $0.97\,\sigma$.

The electrostatic interaction between two charges, $q_i$ and $q_j$, separated by a distance $r_{ij}$, is described by Coulomb's law:
\begin{equation}
    U_{\rm Coul}(r_{ij}) = \frac{q_i q_j}{4 \pi \varepsilon_0 \varepsilon_r r_{ij}}
    \label{eq:Coul}
\end{equation}
where $\varepsilon_0$ is the vacuum permittivity and $\varepsilon_r$ is the relative dielectric constant of the medium.
Electrostatic interactions were evaluated using the standard three-dimensional particle--particle particle--mesh (PPPM) method. 
In this approach, the Coulomb potential is divided into short-range real-space and long-range reciprocal-space contributions. 
A real-space cutoff of $3.0\,\sigma$ is employed, while the reciprocal-space part is computed efficiently on a mesh using Fourier-space techniques.
The ion valence is $z_i=1$, corresponding to monovalent ions. The reduced charge is defined as $q^* = z_i e/\sqrt{4\pi\varepsilon_0\varepsilon_r\sigma\varepsilon}$ and is set to $q^*=1.0$. 
The reduced temperature is taken as $T^* = k_{\mathrm{B}}T/\varepsilon = 1.2$ \cite{kumar2005polyelectrolyte}. Accordingly, the reduced Bjerrum length is expressed as $l_{\mathrm{B}}/\sigma = e^2/(4\pi\varepsilon_0\varepsilon_r k_{\mathrm{B}}T\,\sigma) = (q^*)^2/T^*$, yielding $l_{\mathrm{B}}/\sigma \approx 0.83$. 
This corresponds to a weak-to-moderate electrostatic coupling regime and is comparable in magnitude to aqueous polyelectrolyte systems at room temperature.
Notably, in this implicit-solvent framework, water molecules and explicit hydration shells are not included. 
Therefore, ion size in this CG model should be interpreted as an effective parameter that accounts for both the bare ionic radius and part of the hydration contribution.

The wall potential $U_{\rm wall}$ is applied at the grafting surface at the bottom of the simulation box. 
This potential exerts a short-range repulsive force on all particles when their distance from the wall along the $z$ direction is less than $0.5\,\sigma$. 
We used $U_{\rm wall}$ similar to the shifted LJ potential as defined in Eq.~(\ref{eq:LJ}):
\begin{equation}
    U_{\rm wall}(z) = \left\{ \begin{array}{cl} \displaystyle
         {4\varepsilon \left\lbrack {\left( \frac{\sigma}{z+\Delta z} \right)^{12} - \left( \frac{\sigma}{z+\Delta z} \right)^{6}} \right\rbrack + \varepsilon}
         & , z < 0.5\sigma\\
        {0}& , z \geq 0.5\sigma \end{array} \right.
    \label{eq:wall}
\end{equation}
with $\Delta z = (\sqrt[6]{2} - 0.5)\,\sigma$. 

The time evolution of the system is simulated by solving the Langevin equation:
\begin{equation}
    m \ddot{\mathbf{r}}_i = - \boldsymbol{\nabla}_i U_{\rm total} - \gamma \dot{\mathbf{r}}_i + \mathbf{f}_i
\end{equation}
where $m$ denotes the particle mass, $\mathbf{r}_i$ is the position vector of the $i$th particle, and $\ddot{\mathbf{r}}_i$ represents its acceleration. 
The terms on the right-hand side correspond to distinct force contributions: 
$- \boldsymbol{\nabla}_i U_{\rm total}$ represents the conservative force derived from the total potential energy, 
$- \gamma \dot{\mathbf{r}}_i$ denotes the damping force with friction coefficient $\gamma$, 
and $\mathbf{f}_i$ is a stochastic force term accounting for thermal fluctuations.
The stochastic force satisfies the fluctuation--dissipation theorem: 
\begin{align}
    \langle \mathbf{f}_i(t) \rangle &= 0, \\
    \langle \mathbf{f}_{i}(t)\mathbf{f}_{j}(t') \rangle
    &= 2 k_{\rm B} T \gamma\, \delta_{ij}\, \delta(t - t')
\end{align}
where $k_{\rm B}$ is the Boltzmann constant and $T$ is the system temperature. 

In the simulations, all particles are assigned a unit mass ($m = 1$), and the friction coefficient is set to $\gamma = 0.5\,m/\tau$, where the time unit is defined as $\tau = \sqrt{m\sigma^2/\varepsilon}$. 
The integration time step is $\Delta t = 0.001\,\tau$.
For the salt-containing PE brush model, each system is equilibrated for $6\times10^4\,\tau$ to obtain stable configurations, followed by an additional production run of $5\times10^3\,\tau$ for data collection. 
To improve statistical reliability, three independent simulations are performed for each parameter set, from which error bars are obtained.

\section{Results and Discussions}

\subsection{Decrease $\sigma_{\mathrm{C}}$ and keep $\sigma_{\mathrm{T}} = 1.0$}
\label{sec:PEB_C}

Compared with the point-charge approximation in classical theories, real ions possess finite sizes, which can induce spatial heterogeneity and nonuniform charge distributions. 
In this section, we systematically investigate the effect of counterion size $\sigma_\mathrm{C}$ on the structural and electrostatic properties of PE brushes at different salt concentrations.

\begin{figure}[htbp]
  \centering
  \includegraphics[width=1.0\textwidth]{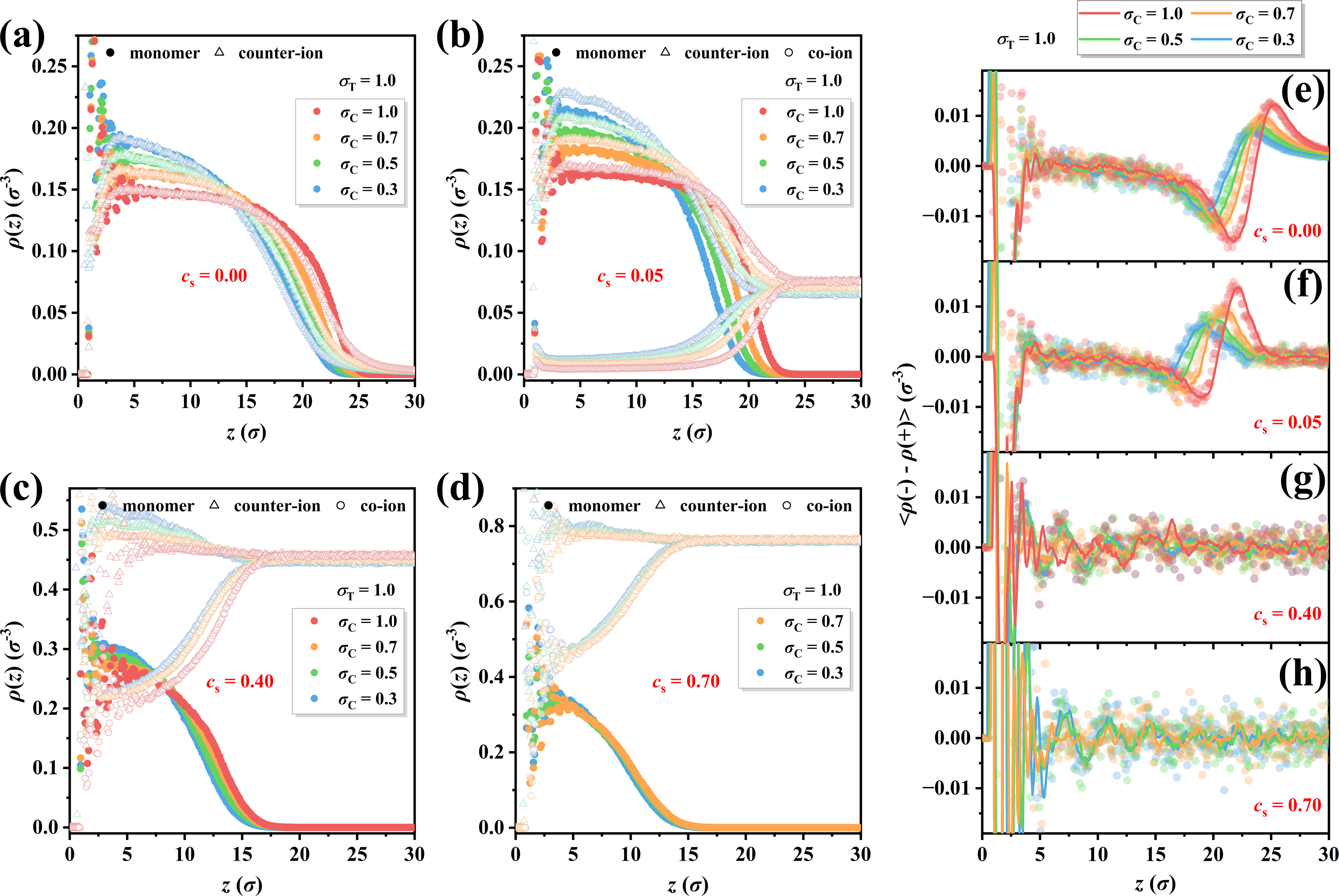}
  \caption{Density profiles of PE monomers (solid circles), counterions (triangles), and co-ions (open circles) along the direction normal to the grafting surface for different counterion size $\sigma_\mathrm{C}$ with fixed co-ion size $\sigma_\mathrm{T}=1.0$ at different salt concentrations: $c_{s}$ = 0 (a), 0.05 (b), 0.40 (c), and 0.70 (d). Panels (e--h) show the corresponding local net charge distributions. Solid lines represent smoothed profiles obtained by convolution, and symbols denote raw data. The red curve for $c_\mathrm{s}=0.70$ is absent because the maximum accessible salt concentration in the reference system is $c_\mathrm{s}=0.45$.}
  \label{fig:C_rho}
\end{figure}

\textbf{Distribution of Monomers, Counterions and Co-ions.}
We first examine the number density profiles $\rho(z)$ along the direction normal to the grafting surface. 
As shown in \textbf{Figure~\ref{fig:C_rho}}a, in the salt-free case the monomer distribution exhibits a step-like profile for $\sigma_\mathrm{C}=1.0$, indicating a stretched brush conformation. 
Decreasing $\sigma_\mathrm{C}$ shifts the monomer density toward smaller $z$ and reduces the distribution width, indicating progressive brush collapse. 
This behavior is consistent with enhanced penetration of smaller counterions into the brush interior.
Counterion distributions closely follow the monomer profiles within the brush, reflecting strong electrostatic attraction. 
Co-ions are largely excluded from the brush but partially penetrate in regions of local charge overcompensation induced by counterion accumulation. 
The corresponding net charge profiles (\textbf{Figure~\ref{fig:C_rho}}e) show that the brush interior remains approximately electroneutral, while interfacial charge oscillations are present and become weaker for smaller counterions, indicating more homogeneous charge screening.
At low salt concentration ($c_\mathrm{s}=0.05$, \textbf{Figure~\ref{fig:C_rho}}b), the brush further collapses, with increased monomer density near the grafting surface. Counterions are strongly enriched inside the brush, while co-ions remain mostly excluded but partially penetrate due to local overcompensation. 
The net charge oscillations at the interface are significantly reduced compared with the salt-free case, especially for smaller counterions, indicating more uniform screening.
At higher salt concentrations ($c_\mathrm{s}=0.40$ and $0.70$, \textbf{Figure~\ref{fig:C_rho}}c,d), the brush collapses further and both counterions and co-ions penetrate deeper into the brush. 
The density profiles for different $\sigma_\mathrm{C}$ values gradually converge, indicating that strong electrostatic screening dominates the system behavior. 
Consistently, interfacial charge oscillations weaken, while internal charge structuring becomes more pronounced under high-salt conditions.
Taken together, decreasing counterion size enhances ion penetration and local accumulation, promoting charge screening and brush collapse. 
This size effect is most significant at low salt concentration but becomes progressively suppressed as electrostatic screening dominates at high salt concentration.

\textbf{Brush Height and End-monomer Distribution.}
To quantify the brush structure, we calculate the brush height $H$, defined as:
\begin{equation}
  H=\frac{2\int_0^{L_z} z \rho_{\mathrm{m}}(z) \mathrm{d} z}{\int_0^{L_z} \rho_{\mathrm{m}}(z) \mathrm{d} z}.
  \label{eq:PEB_H}
\end{equation}
Here, $z$ is the distance of a PE monomer from the grafting surface, and $\rho_{\mathrm{m}}(z)$ represents the monomer number density profile along $z$. 
The prefactor $2$ originates from the relation between the first moment of the density distribution and the brush height. 
Assuming a nearly uniform density within the brush, the brush height can be approximated as twice the first moment \cite{ibergay2010mesoscale}. 
The corresponding values are summarized in \textbf{Tables~S1--S3}.

\begin{figure}[htbp]
  \centering
  \includegraphics[width=0.67\textwidth]{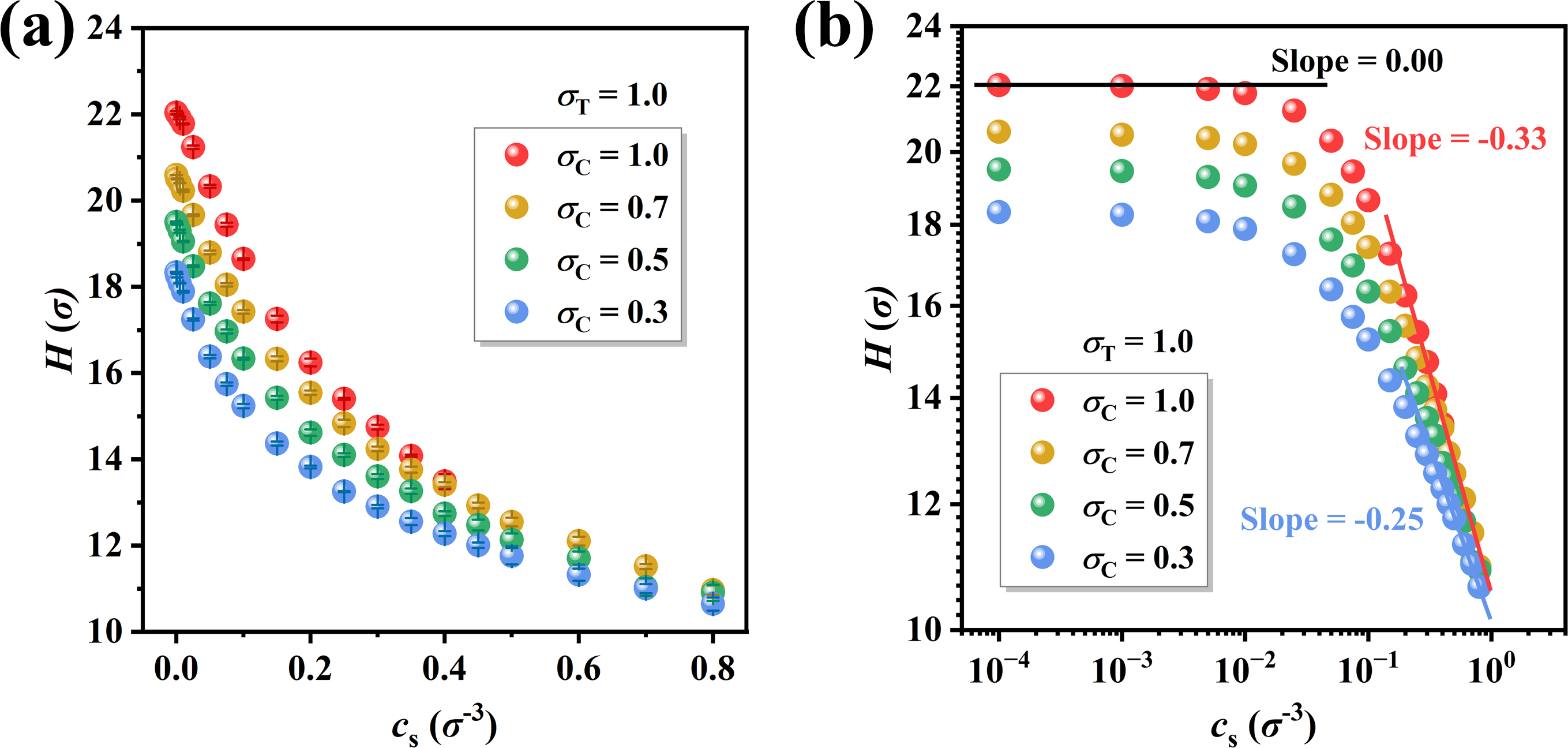}
    \caption{PE brush height $H$ as a function of salt concentration $c_\mathrm{s}$ for different $\sigma_\mathrm{C}$ at $\sigma_\mathrm{T}=1.0$: (a) linear scale; (b) log--log scale.}
  \label{fig:C_H}
\end{figure}

As shown in \textbf{Figure~\ref{fig:C_H}}a, the brush height decreases monotonically with increasing salt concentration for all counterion sizes, indicating progressive brush collapse. 
Smaller counterions consistently lead to lower brush heights, suggesting enhanced penetration into the brush and stronger electrostatic screening, which reduces monomer repulsion and promotes chain contraction. 
The salt dependence is strong in the low-salt regime and becomes progressively weaker at high salt concentration, where the differences between ion sizes diminish.
To examine the scaling behavior, the data are plotted on a log--log scale in \textbf{Figure~\ref{fig:C_H}}b. 
When $c_\mathrm{s}\lesssim10^{-2}$, the brush height is nearly independent of salt concentration ($H \propto c_\mathrm{s}^{0}$), corresponding to the osmotic brush regime. 
At higher salt concentrations, the system enters the salted brush regime, where $H$ follows a power-law dependence on $c_\mathrm{s}$. 
For $\sigma_{\mathrm{C}}=1.0$, the scaling exponent is $\alpha\approx-0.33$, consistent with the classical prediction of $-1/3$. 
In contrast, for smaller counterions ($\sigma_{\mathrm{C}}=0.3$), the exponent shifts to $\alpha\approx-0.25$, indicating a deviation from ideal scaling due to stronger ion penetration and heterogeneous screening. 
A crossover regime is observed between these limits, where osmotic and screening effects compete and no single power law applies. 
Overall, ion size significantly modifies the effective scaling behavior of the brush height.

For ideal linear polymer chains, the chain-end distribution typically follows a Gaussian profile. 
In PE brush systems, electrostatic interactions lead to significant deviations from this ideal behavior. 
As shown in \textbf{Figure~S3}, the chain-end distributions for $\sigma_\mathrm{T}=\sigma_\mathrm{C}=1.0$ remain unimodal across all salt concentrations, with the peak shifting toward the grafting surface as salt concentration increases, indicating progressive brush contraction.
To quantify these conformational changes, three descriptors are extracted from the $z$-direction chain-end distribution: the peak position $z_{\max}$, peak height $\rho_{\max}$, and full width at half maximum (FWHM), representing the average extension, local accumulation, and distribution width, respectively.

\begin{figure}[htbp]
   \centering
    \includegraphics[width=1.0\textwidth]{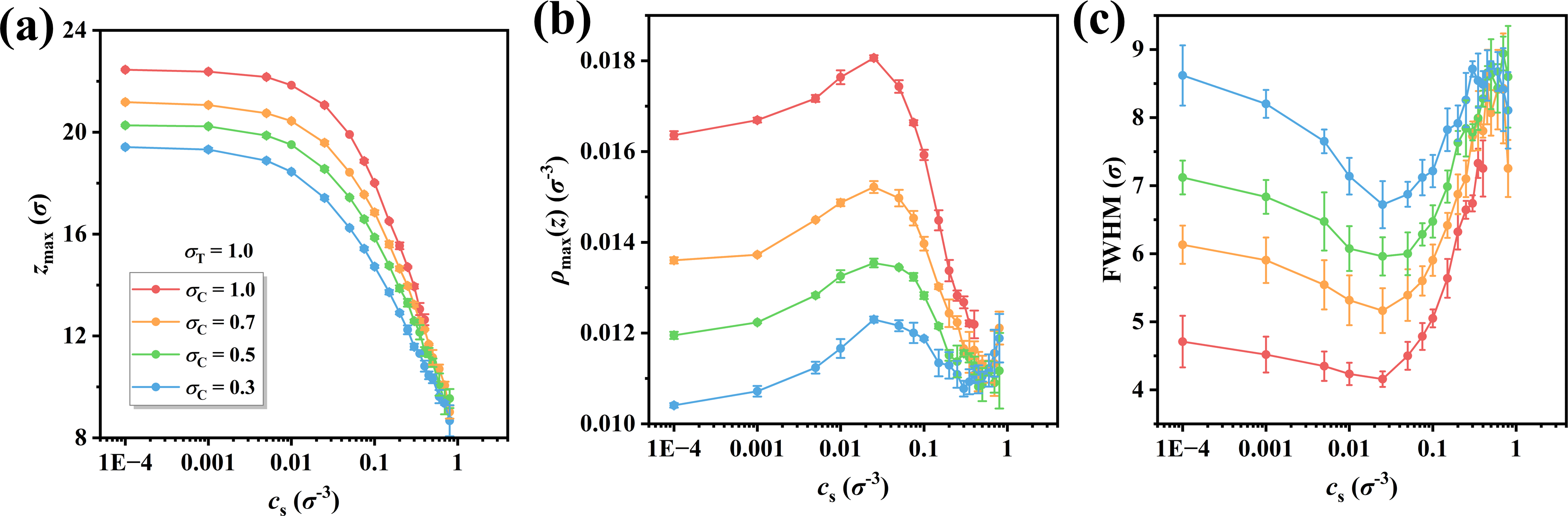}
    \caption{Dependence of characteristic parameters of the PE chain-end monomer density distribution on salt concentration $c_\mathrm{s}$ for different $\sigma_\mathrm{C}$ at $\sigma_\mathrm{T}=1.0$: (a) peak position $z_{\max}$, (b) peak height $\rho_{\max}$, and (c) full width at half maximum (FWHM).}
   \label{fig:C_end_distribution}
\end{figure}

As shown in \textbf{Figure~\ref{fig:C_end_distribution}}a, $z_{\max}$ decreases monotonically with increasing salt concentration, consistent with the reduction in brush height $H$. 
Smaller counterions lead to lower $z_{\max}$ values, suggesting enhanced penetration into the brush and stronger electrostatic screening, which promotes chain contraction.
In contrast, $\rho_{\max}$ and the FWHM exhibit nonmonotonic dependences on salt concentration (\textbf{Figure~\ref{fig:C_end_distribution}}b and \ref{fig:C_end_distribution}c). 
In the low-salt regime, $\rho_{\max}$ increases while the FWHM decreases, indicating a more localized chain-end distribution. 
With further increase in salt concentration, both quantities show a crossover near $c_\mathrm{s}\approx0.025$, where $\rho_{\max}$ decreases and the FWHM increases. 
This trend indicates a transition toward a more dispersed chain-end distribution.
As the counterion size decreases, $\rho_{\max}$ decreases and the FWHM increases, indicating broader chain conformations and reduced localization. 
At high salt concentration, the differences between different ion sizes become weak, and the values of $\rho_{\max}$ and FWHM gradually converge. 
This convergence suggests that strong electrostatic screening suppresses long-range interactions, making chain conformations primarily governed by short-range interactions and excluded-volume effects.

\textbf{Pair Correlation Analysis between PE Monomers and Counterions/Co-ions.}
To gain insight into the underlying interaction mechanisms, we analysis the radial distribution function (RDF) of counterions and co-ions relative to PE monomers. 
For two distinct species A and B, the RDF is defined as:
\begin{equation}
g_{\mathrm{AB}}(r) = \frac{1}{4\pi r^2 \Delta r \, \rho_{\mathrm{B}}} \cdot \frac{1}{N_{\mathrm{A}}} 
\left\langle \sum_{i=1}^{N_{\mathrm{A}}} n_i^{\mathrm{AB}}(r, r+\Delta r) \right\rangle
\end{equation}
where $\rho_{\mathrm{B}} = N_{\mathrm{B}} / V$ represents the average number density of species $B$, and $n_i^{\mathrm{AB}}(r, r+\Delta r)$ denotes the number of $B$ particles located within the spherical shell $(r, r+\Delta r)$ centered on the $i$-th particle of species $A$. 
In this context, $N_{\mathrm{A}}$ and $N_{\mathrm{B}}$ are the total numbers of particles of species $A$ and $B$, respectively, and $V$ is the volume of the simulation box.

\begin{figure}[htbp]
   \centering
    \includegraphics[width=1.0\textwidth]{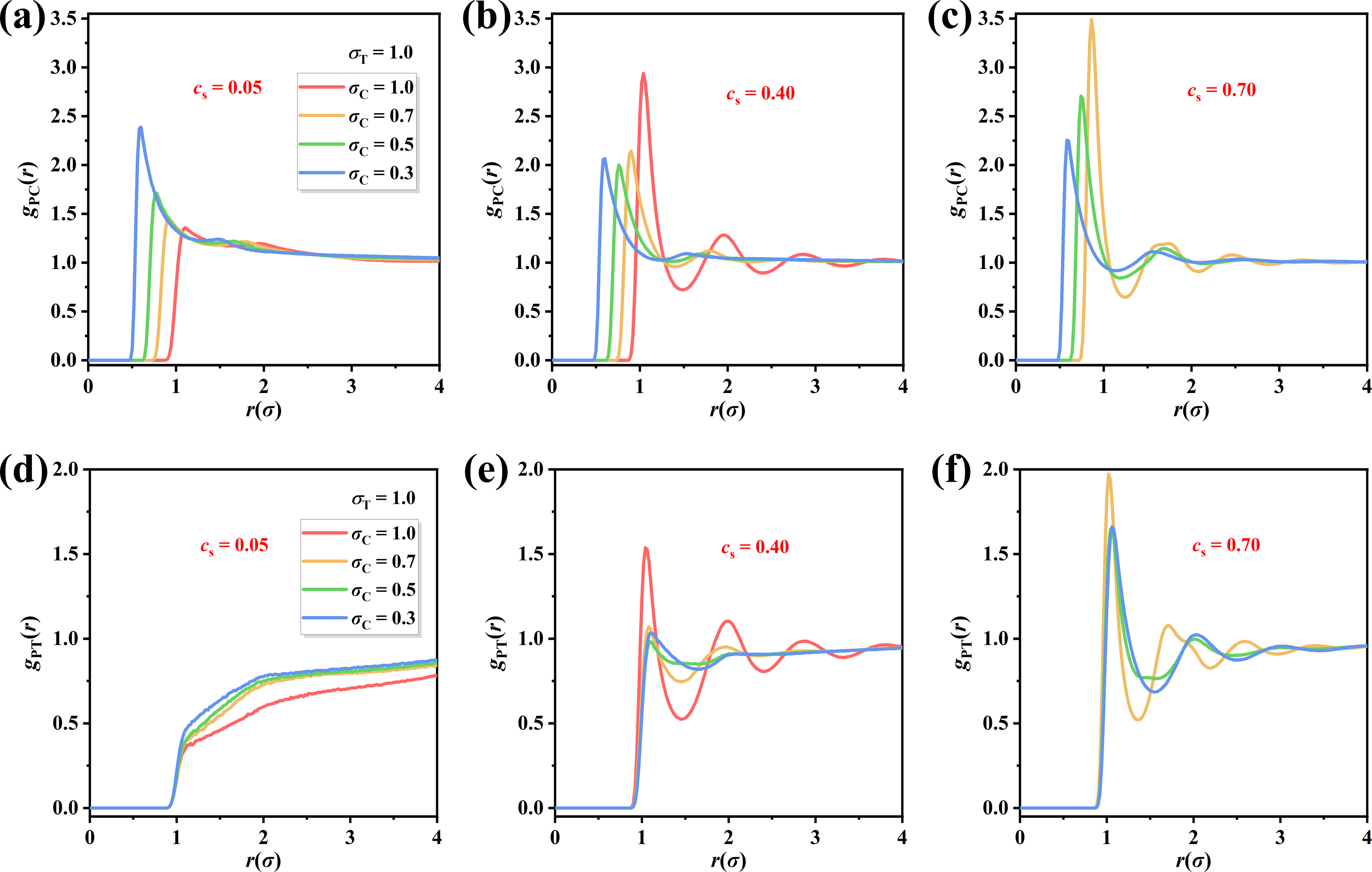}
    \caption{Radial distribution functions (RDFs) between PE monomers and counterions (a--c) or co-ions (d--f) for different $\sigma_\mathrm{C}$ at $\sigma_\mathrm{T}=1.0$ under various salt concentrations. The red curve for $c_\mathrm{s}=0.70$ is absent because the maximum accessible salt concentration in the reference system is $c_\mathrm{s}=0.45$.}
   \label{fig:C_rdf}
\end{figure}

As shown in \textbf{Figure~\ref{fig:C_rdf}}a--c, all $g_{\mathrm{PC}}(r)$ profiles exhibit a pronounced first-neighbor peak, indicating strong short-range attraction between PE monomers and counterions. 
At low salt concentrations ($c_\mathrm{s}=0.00$ and $0.05$, \textbf{Figure~S4} and \textbf{Figure~\ref{fig:C_rdf}}a), 
decreasing the counterion size $\sigma_\mathrm{C}$ shifts the first peak toward smaller $r$ and enhances its intensity, indicating that smaller counterions approach PE monomers more closely.
As $c_\mathrm{s}=0.40$ (\textbf{Figure~\ref{fig:C_rdf}}b), brush collapse increases the local ion density and enhances the first peak for most counterion sizes. 
For large counterions ($\sigma_\mathrm{C}=1.0$), additional second and third peaks emerge, indicating the formation of layered coordination structures. 
In contrast, the smallest counterions ($\sigma_\mathrm{C}=0.3$) exhibit a reduced first peak, suggesting competing effects of spatial confinement and ion competition.
At high salt concentration ($c_\mathrm{s}=0.70$, \textbf{Figure~\ref{fig:C_rdf}}c), the first peak remains enhanced for all counterion sizes, while the differences between different $\sigma_\mathrm{C}$ values become less pronounced because of strong electrostatic screening. 
Notably, the smallest counterions show the weakest peak enhancement, leading to an inversion in the size dependence of the peak height.

\begin{figure}[htbp]
   \centering
    \includegraphics[width=1.0\textwidth]{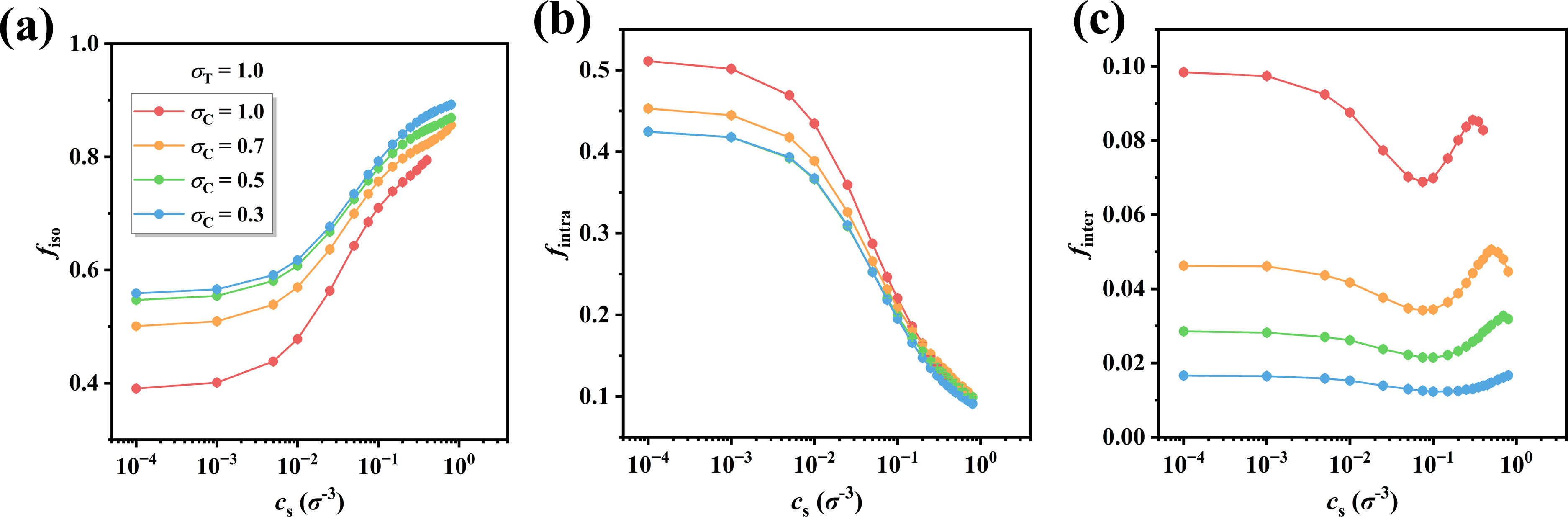}
    \caption{Fractions of different counterion states as functions of salt concentration for different $\sigma_\mathrm{C}$ at $\sigma_\mathrm{T}=1.0$: (a) isolated ($f_{\mathrm{iso}}$), (b) intrachain condensation ($f_{\mathrm{intra}}$), and (c) interchain bridging ($f_{\mathrm{inter}}$). Error bars are smaller than the symbol size.}
   \label{fig:C_counterion}
\end{figure}

To characterize the bridging behavior between counterions and PE chains, counterions are classified into three states following Miao \emph{et al.} \cite{tan2023size}: isolated ($f_{\mathrm{iso}}$), intrachain condensation ($f_{\mathrm{intra}}$), and interchain bridging ($f_{\mathrm{inter}}$). 
A cutoff distance $R_{\mathrm{c}}=\sqrt{2}(D_{\mathrm{P}}+D_{\mathrm{C}})/2$ is used to identify condensed counterions \cite{liu2017heterogeneous, hao2020surface, hao2025systematic}. 
This cutoff is slightly larger than the direct contact distance $(D_{\mathrm{P}}+D_{\mathrm{C}})/2$ and approximately corresponds to the end of the first RDF peak.
Condensed counterions associated with monomers on the same chain are classified as intrachain condensation, whereas those associated with different chains are identified as interchain bridging.
As shown in \textbf{Figure~\ref{fig:C_counterion}}a, the fraction of isolated counterions increases with increasing salt concentration, and smaller counterions consistently exhibit higher free fractions. 
In contrast, the probability of intrachain condensation decreases with increasing salt concentration (\textbf{Figure~\ref{fig:C_counterion}}b). Under low-salt conditions, smaller counterions show weaker intrachain condensation, while the size dependence becomes less pronounced at high salt concentrations.
The fraction of interchain bridging is generally very low for monovalent counterions (\textbf{Figure~\ref{fig:C_counterion}}c). 
Nevertheless, decreasing counterion size further suppresses interchain bridging. 
Overall, although smaller counterions approach PE monomers more closely, they are less likely to simultaneously coordinate multiple monomers or chains. As a result, the isolated fraction increases, whereas both intrachain condensation and interchain bridging are suppressed.

Since co-ions carry the same charge as PE monomers, their distribution around the chains is governed by the competition between electrostatic repulsion from PE monomers and attractive interactions with counterions. 
As shown in \textbf{Figure~\ref{fig:C_rdf}}d, at low salt concentration ($c_\mathrm{s}= 0.05$), the P--T correlation remains below unity in the short-range region, indicating that co-ions are largely excluded from the vicinity of PE monomers. 
Smaller counterions exhibit stronger accumulation near the monomers, leading to a slightly enhanced co-ion presence compared with the case of larger counterions ($\sigma_\mathrm{C}=1.0$), although repulsion still dominates the overall behavior.
As the salt concentration increases to $c_\mathrm{s}= 0.40$ (\textbf{Figure~\ref{fig:C_rdf}}e), a pronounced first peak appears in the P--T correlation function, while larger counterions ($\sigma_\mathrm{C}=1.0$) also induce visible second and third peaks. 
This result indicates that counterion accumulation near PE monomers leads to local charge overcompensation, allowing part of the co-ions to enter the first coordination shell and participate in local charge neutralization. 
In contrast, for $\sigma_\mathrm{C}=0.7$, $0.5$, and $0.3$, the weaker higher-order peaks suggest that co-ions remain less accessible to the near-monomer region.
At high salt concentration ($c_\mathrm{s}= 0.70$, \textbf{Figure~\ref{fig:C_rdf}}f), the P--T correlations become further enhanced and the coordination structures more pronounced. 
For $\sigma_\mathrm{C}=0.7$, the first peak reaches its maximum and the higher-order peaks shift toward smaller $r$, reflecting the formation of a more compact locally electroneutral structure. 
By contrast, for $\sigma_\mathrm{C}=0.5$ and $0.3$, the peaks become weaker and shift outward, indicating a more pronounced charge-layered structure with oscillatory charge distributions inside the brush.
Overall, the P--C and P--T correlations evolve cooperatively with salt concentration and ion size: counterions primarily determine the local coordination structures, while co-ions participate in charge compensation, jointly regulating the electrostatic environment within the brush.

In summary, counterion size and salt concentration jointly regulate the structural behavior of PE brushes. 
Small counterions exhibit strong local coordination across the entire salt range, resulting in a weak dependence of brush structure on salt concentration. 
In contrast, larger counterions are restricted by excluded-volume effects at low salt concentration, while this limitation is partially mitigated at high salt concentration due to enhanced screening and ion crowding.
For the scaling behavior, all systems exhibit an osmotic brush regime at low salt concentration with $\alpha \approx 0$. 
At higher salt concentrations, the reference system ($\sigma_\mathrm{C}=1.0$) recovers the classical salted brush scaling with $\alpha \approx -1/3$, whereas decreasing counterion size leads to a deviation from this scaling, with the exponent shifting to $\alpha \approx -0.25$.
Overall, PE brush behavior is governed by the interplay between ion penetration and electrostatic screening, while ion-size effects become progressively weaker in the high-salt regime.

\subsection{Decrease $\sigma_{\mathrm{T}}$ and keep $\sigma_{\mathrm{C}} = 1.0$}
\label{sec:PEB_T}

We next examine the effect of co-ion size on PE brush by fixing the counterion size at $\sigma_{\mathrm{C}}=1.0$ and varying the co-ion size $\sigma_{\mathrm{T}}$.
The salt-free case is not considered in this and the following sections because co-ions are absent and this regime has already been discussed for varying $\sigma_{\mathrm{C}}$. 

\begin{figure}[htbp]
   \centering
   \includegraphics[width=1.0\textwidth]{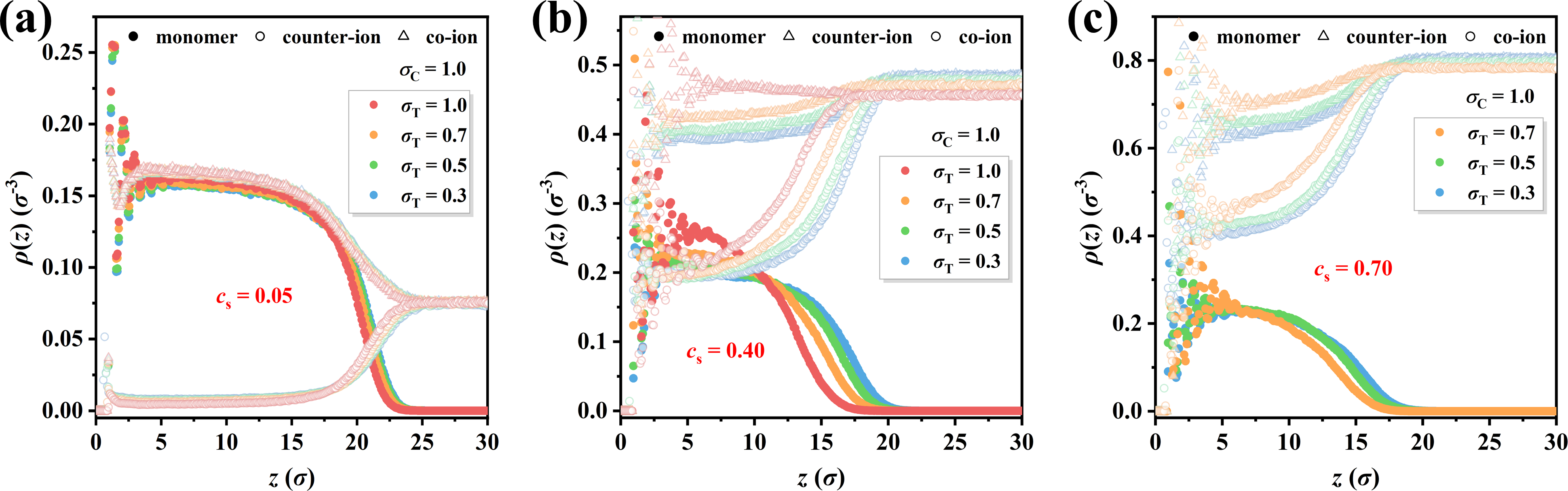}
    \caption{Density profiles of PE monomers, counterions, and co-ions along the direction normal to the grafting surface for different co-ion size $\sigma_{\mathrm{T}}$ with fixed counterion size $\sigma_\mathrm{C}=1.0$ at different salt concentrations: $c_\mathrm{s}=0.05$ (a), $0.40$ (b), and $0.70$ (c). The red curve for $c_\mathrm{s}=0.70$ is absent because the maximum accessible salt concentration in the reference system is $c_\mathrm{s}=0.45$.}
   \label{fig:T_rho}
\end{figure}

\textbf{Distribution of Monomers, Counterions and Co-ions.}
We first focus on the density distributions of the three types of particles.
At low salt concentration ($c_\mathrm{s} = 0.05$, \textbf{Figure~\ref{fig:T_rho}}a), co-ions remain largely excluded from the brush, and varying $\sigma_{\mathrm{T}}$ produces only minor changes in the internal density profiles. 
Smaller co-ions slightly enhance interfacial charge oscillations (\textbf{Figure~S5}a).
At intermediate and high salt concentrations ($c_\mathrm{s} = 0.40$ and $0.70$, \textbf{Figure~\ref{fig:T_rho}}b and \ref{fig:T_rho}c), the co-ion size effect becomes more pronounced. 
Smaller co-ions show weaker accumulation within the brush, leading to an outward shift of the brush--solution interface and a more extended brush structure. 
Correspondingly, the local charge distributions (\textbf{Figure~S5}b and \textbf{S5}c) exhibit stronger charge oscillations and weaker local electroneutrality for smaller $\sigma_{\mathrm{T}}$.
These results indicate that reducing the co-ion size suppresses ion condensation and local charge compensation within the brush, thereby promoting brush swelling.

\begin{figure}[htbp]
    \centering
    \includegraphics[width=0.67\textwidth]{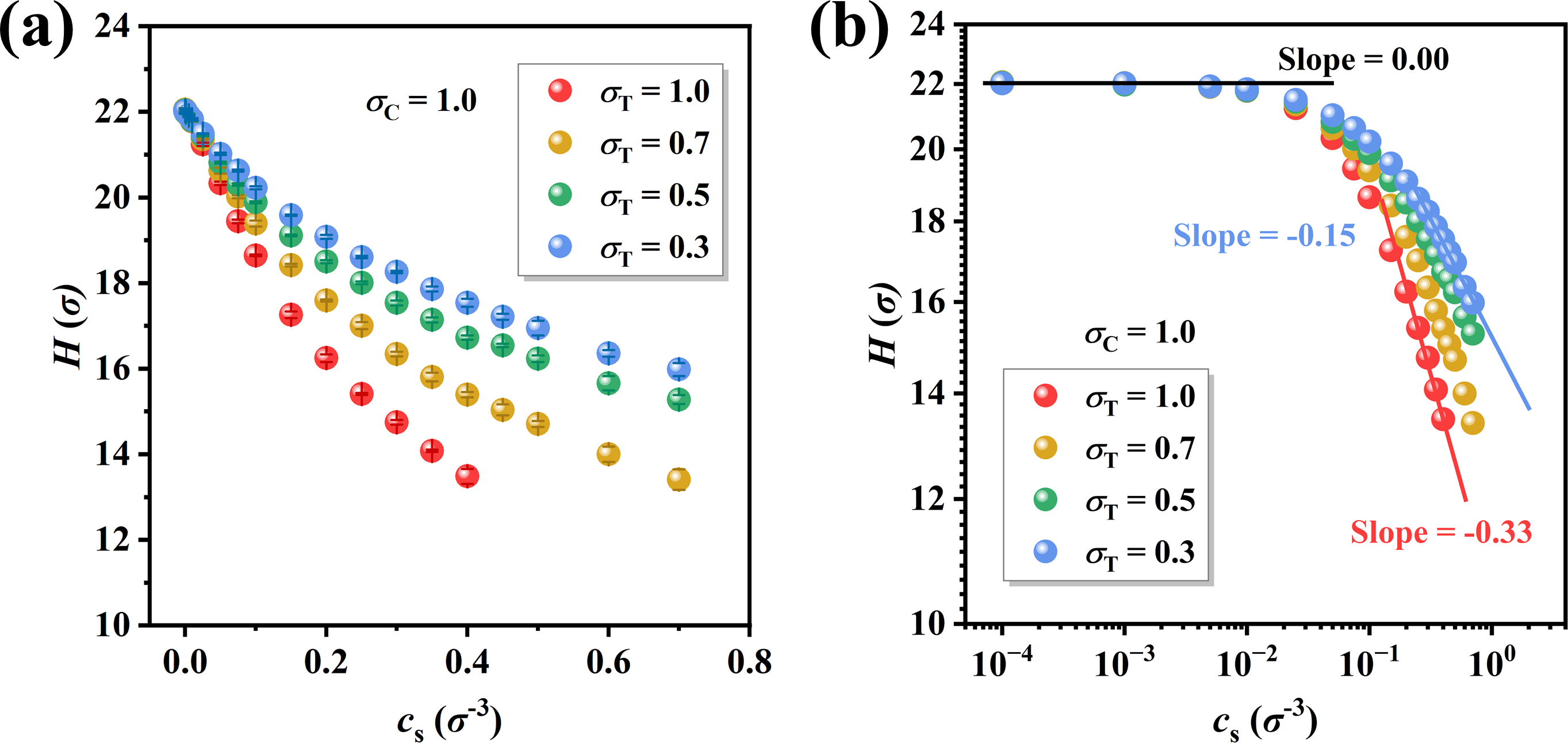}
    \caption{PE brush height $H$ as a function of salt concentration $c_\mathrm{s}$ for different $\sigma_\mathrm{T}$ at $\sigma_\mathrm{C} = 1.0$: (a) linear scale; (b) log--log scale.}
    \label{fig:T_H}
\end{figure}

\textbf{Brush Height and End-monomer Distribution.}
As shown in \textbf{Figure~\ref{fig:T_H}}a, the brush height decreases monotonically with increasing salt concentration, and the effect of co-ion size becomes more pronounced at high salt concentrations.
Smaller co-ions lead to a weaker reduction in brush height, indicating a reduced degree of brush collapse.
The log--log representation (\textbf{Figure~\ref{fig:T_H}}b) shows that in the low-salt regime ($c_\mathrm{s} \lesssim 10^{-2}$), all curves nearly overlap, suggesting  a negligible effect of co-ion size in the osmotic brush regime. 
At higher salt concentrations, the system enters a salted brush regime where the scaling exponent decreases with decreasing $\sigma_{\mathrm{T}}$. 
For $\sigma_{\mathrm{T}} = 0.3$, the exponent reaches $\alpha \approx -0.14$, reflecting a weakened response of the brush to salt concentration.

As shown in \textbf{Figure~S7}a, $z_{\max}$ decreases monotonically with increasing salt concentration.
The reduction in $z_{\max}$ becomes weaker as the co-ion size decreases, particularly at intermediate and high salt concentrations.
The peak height and distribution width (\textbf{Figure~S7}b and \textbf{S7}c) exhibit similar trends. 
In the low-salt regime ($c_\mathrm{s} \lesssim 0.025$), $\rho_{\max}$ and FWHM nearly overlap for different co-ion sizes, consistent with the trend of brush height $H$. 
At higher salt concentrations, smaller co-ions lead to a slower decrease in $\rho_{\max}$ and a weaker increase in FWHM, indicating suppressed inward shift and broadening of the chain-end distributions. 

\begin{figure}[htbp]
   \centering
    \includegraphics[width=1.0\textwidth]{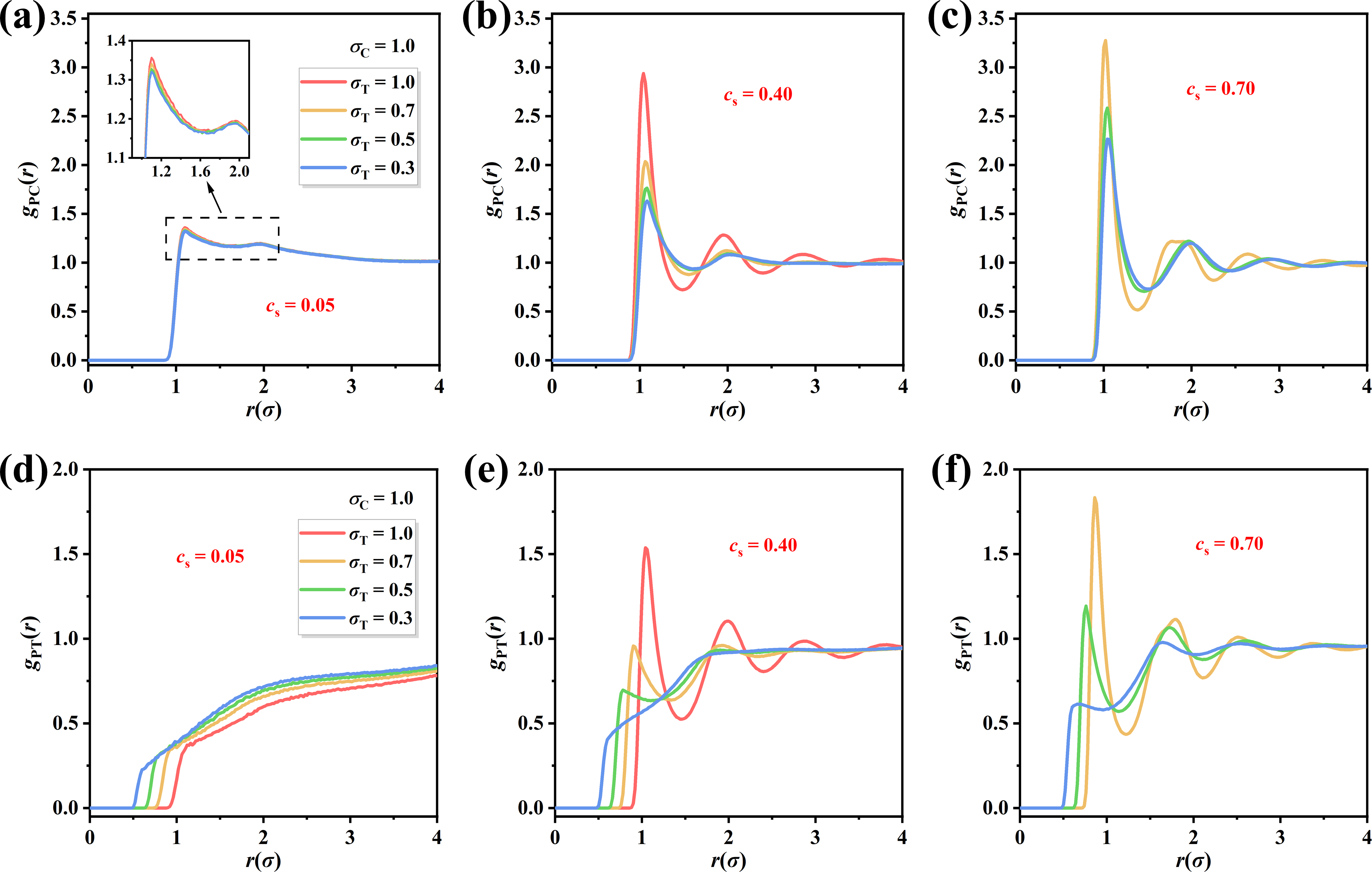}
    \caption{RDFs between PE monomers and counterions (a--c) or co-ions (d--f) for different $\sigma_\mathrm{T}$ at $\sigma_\mathrm{C}=1.0$ under various salt concentrations. The red curve for $c_\mathrm{s}=0.70$ is absent because the maximum accessible salt concentration in the reference system is $c_\mathrm{s}=0.45$.}
   \label{fig:T_rdf}
\end{figure}

\textbf{Pair Correlation Analysis between PE Monomers and Counterions/Co-ions.}
Co-ions are largely excluded from the brush due to electrostatic repulsion, yet they indirectly influence P--C coordination by modifying the local electrostatic environment. 
Therefore, the radial distribution function $g_{\mathrm{PC}}(r)$ is analyzed at different salt concentrations and co-ion sizes.
As shown in \textbf{Figure~\ref{fig:T_rdf}}a, $g_{\mathrm{PC}}(r)$ profiles are nearly identical for different $\sigma_{\mathrm{T}}$ at low salt concentration ($c_\mathrm{s} = 0.05$), indicating a negligible co-ion size effects. 
Under such conditions, counterion distribution is mainly dictated by direct electrostatic attraction to PE monomers.
When $c_\mathrm{s} = 0.40$ (\textbf{Figure~\ref{fig:T_rdf}}b), a clear dependence on co-ion size emerges: larger co-ions exhibit pronounced multi-peak structures, whereas these features are weakened for smaller $\sigma_{\mathrm{T}}$, reflecting reduced confinement and weaker P--C coordination.
At higher salt concentration ($c_\mathrm{s} = 0.70$, \textbf{Figure~\ref{fig:T_rdf}}c), multi-peak structures reappear for all co-ion sizes with enhanced peak intensities due to ionic crowding. 
In addition, a noticeable inward shift of higher-order peaks for smaller co-ions indicates enhanced ion penetration into the brush and a reorganization of coordination shells.
As shown in \textbf{Figure~S8}, under low-salt conditions the size of co-ions has little effect on the bridging behavior between counterions and PE monomers. 
With increasing salt concentration, however, the interchain bridging ability of counterions is significantly suppressed in the presence of small co-ions.
These results indicate that co-ions regulate P--C coordination and bridging indirectly by modulating the local electrostatic environment, with their size effect becoming more pronounced at high salt concentrations.

To elucidate the role of co-ion size, we analyze the P--T radial distribution function $g_{\mathrm{PT}}(r)$ as a function of salt concentration $c_\mathrm{s}$.
In the low-salt regime ($c_\mathrm{s} = 0.05$, \textbf{Figure~\ref{fig:T_rdf}}d), $g_{\mathrm{PT}}(r)$ shows weak dependence on co-ion size.
The first peak slightly shifts toward smaller $r$ with decreasing $\sigma_{\mathrm{T}}$, while $g_{\mathrm{PT}}(r)<1$ near PE monomers indicates overall depletion of co-ions from coordination shells.
At $c_\mathrm{s} = 0.40$ (\textbf{Figure~\ref{fig:T_rdf}}e), a clear size dependence emerges. 
Smaller co-ions exhibit weakened first coordination peaks and suppressed higher-order structures, indicating reduced participation in local coordination and charge compensation.
When the salt concentration increases to $c_\mathrm{s} = 0.70$ (\textbf{Figure~\ref{fig:T_rdf}}f), P--T correlations are significantly enhanced for all systems, although size dependence remains. 
For $\sigma_{\mathrm{T}}=0.7$, distinct first and second coordination peaks are observed, indicating that co-ions actively participate in organizing the brush interior. As $\sigma_{\mathrm{T}}$ decreases to $0.5$ and $0.3$, the coordination peaks become progressively weaker, and only faint P--T correlations remain at the smallest co-ion size.
These results show that although P--T correlations are generally enhanced at high salt concentration, smaller co-ions are less effectively confined within the brush, resulting in incomplete local charge compensation, consistent with the charge oscillations observed in \textbf{Figure~S5}c.

Based on the above analysis, co-ion size plays an important role in regulating ion distribution and the conformational behavior of PE brushes. 
Small co-ions exhibit weak confinement within the brush, resulting in reduced local coordination and insufficient charge compensation, which suppresses chain collapse. 
In contrast, higher salt concentration enhances counterion accumulation and partially restores P--C coordination within the brush.
The competition between these effects determines the internal ionic structure and ultimately governs the conformational response and scaling behavior of the PE brush. 
Moreover, the deviation from classical scaling laws become more pronounced at small co-ion sizes, reflecting the breakdown of effective local charge compensation.

\subsection{Decrease  $\sigma_\mathrm{C}$ and $\sigma_\mathrm{T}$ simultaneously and keep $\sigma_\mathrm{C} = \sigma_\mathrm{T}$}
\label{sec:PEB_CT}

The effects of counterion and co-ion sizes on PE brush conformations have been discussed separately in the previous sections. Here, we further consider the case where both ion sizes are reduced simultaneously.

\begin{figure}[htbp]
   \centering
    \includegraphics[width=1.0\textwidth]{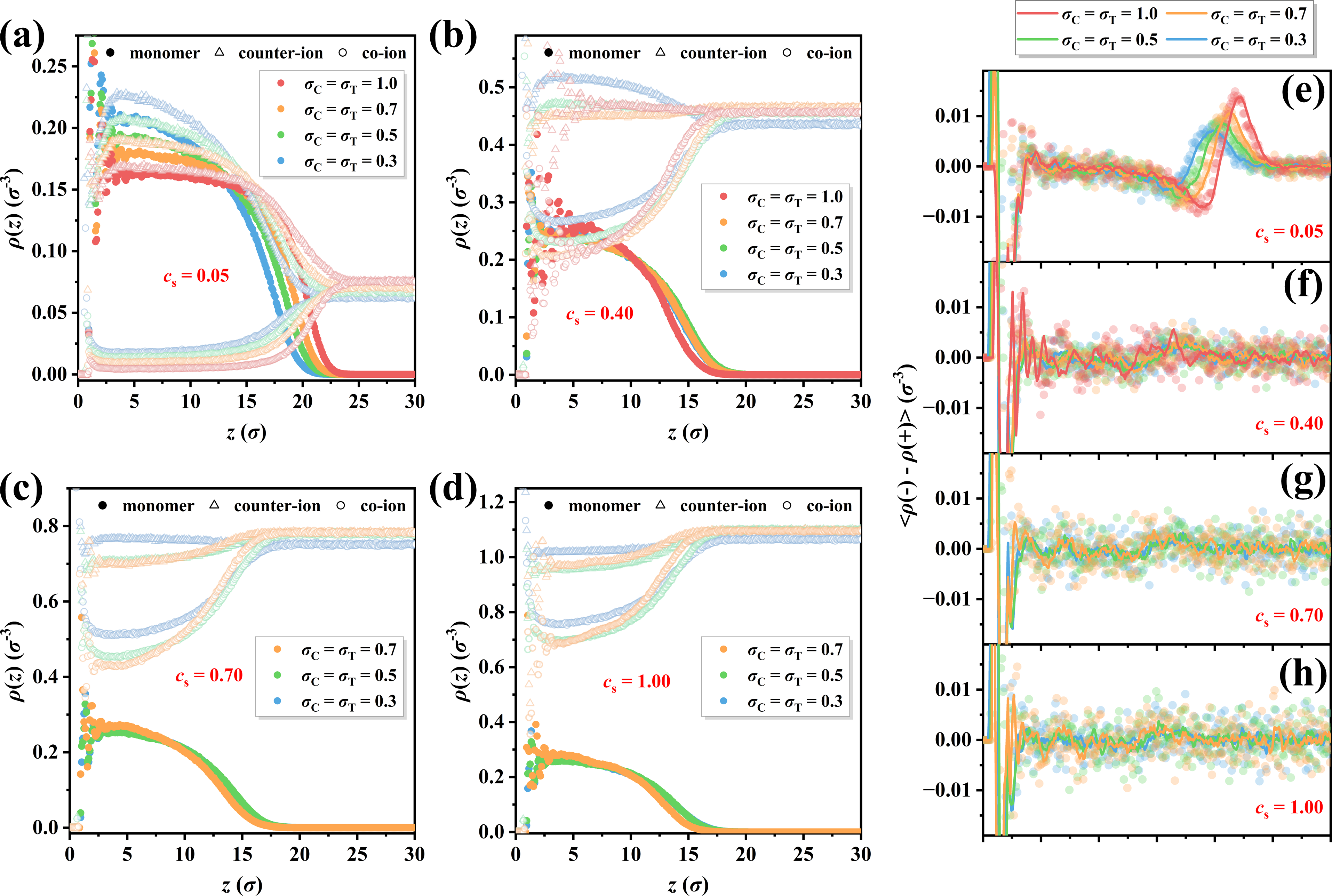}
    \caption{Density profiles of PE monomers, counterions, and co-ions along the direction normal to the grafting surface when counterions and co-ions decrease simultaneously ($\sigma_\mathrm{C} = \sigma_\mathrm{T}$) at different salt concentrations: $c_\mathrm{s}=0.05$ (a), $0.40$ (b), $0.70$ (c), and $1.00$ (d). Panels (e--h) show the corresponding local net charge distributions. Solid lines represent smoothed profiles obtained by convolution, and symbols denote raw data. The red curve for $c_\mathrm{s}=0.70$ is absent because the maximum accessible salt concentration in the reference system is $c_\mathrm{s}=0.45$.}
    \label{fig:CT_rho}
\end{figure}

\textbf{Distribution of Monomers, Counterions and Co-ions.}
We first compare the density profiles at different salt concentrations. 
At low salt concentration ($c_\mathrm{s}=0.05$), as shown in \textbf{Figure~\ref{fig:CT_rho}}a, decreasing the ion size increases the distributions of PE monomers, counterions, and co-ions within the brush region. Meanwhile, the distribution tails near the brush boundary shift toward smaller $z$ values. 
The corresponding net charge profiles (\textbf{Figure~\ref{fig:CT_rho}}e) indicate improved charge balance near the brush interface, with reduced overcompensation and depletion compared to the single-ion-size variation cases.
As the salt concentration increases (\textbf{Figure~\ref{fig:CT_rho}}b--d), ion redistribution becomes more pronounced. 
Both counterions and co-ions increasingly penetrate into the brush, while the monomer distribution shifts toward the grafting surface. 
The sensitivity of the monomer profile to ion size gradually decreases with increasing salt concentration. 
Consistently, the net charge distributions (\textbf{Figure~\ref{fig:CT_rho}}f--h) show weakened interfacial oscillations and more efficient local charge neutralization under high-salt conditions.
Overall, simultaneous reduction of both ion sizes enhances the cooperative penetration of ions into the brush, promoting local charge compensation and driving brush contraction toward the grafting surface. 
Compared with the case of reducing only co-ion size, where ions are more effectively excluded from the brush, the present system allows both ion species to participate in local charge regulation.

\begin{figure}[htbp]
   \centering
    \includegraphics[width=0.67\textwidth]{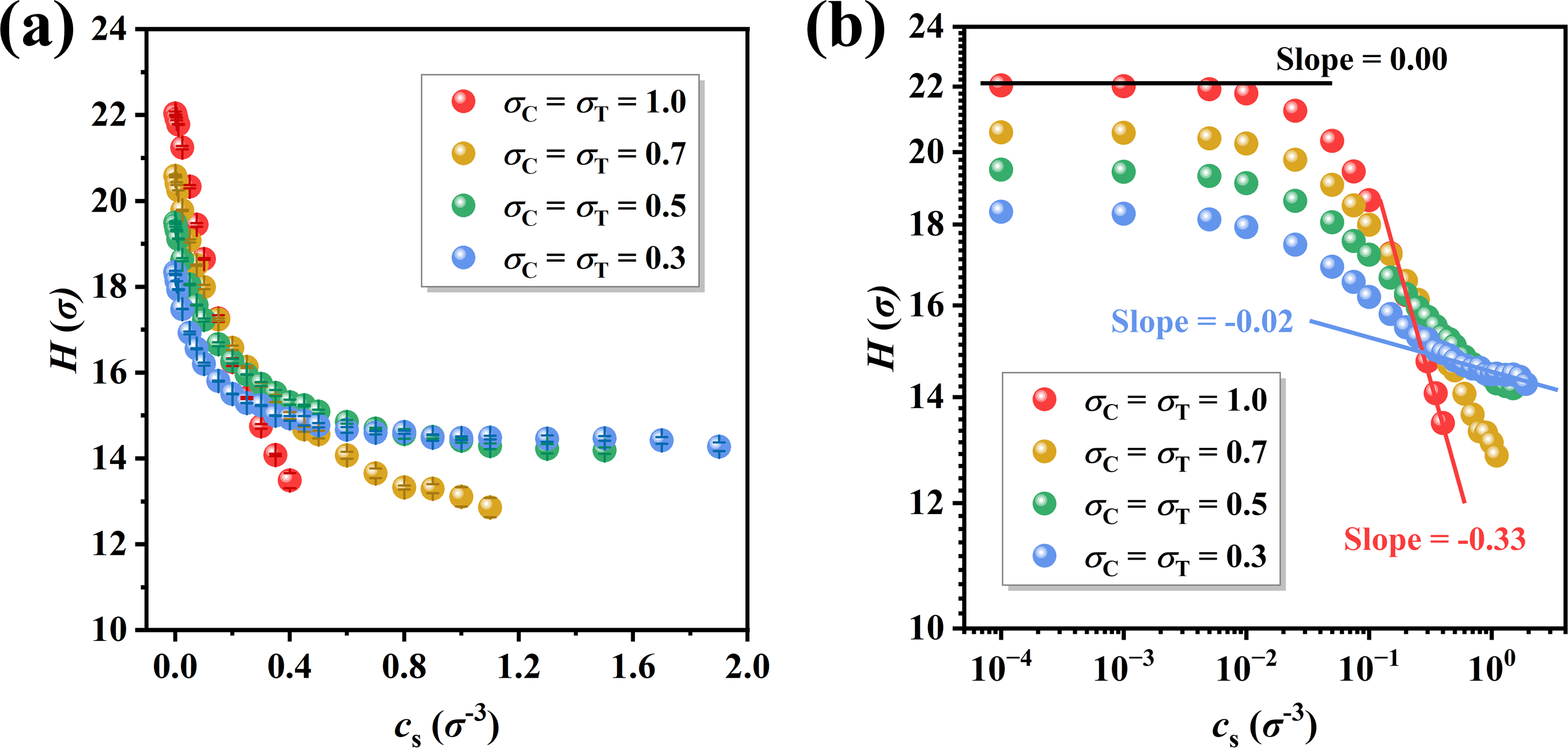}
    \caption{PE brush height $H$ as a function of salt concentration $c_\mathrm{s}$ when counterions and co-ions decrease simultaneously ($\sigma_\mathrm{C} = \sigma_\mathrm{T}$): (a) linear scale; (b) log--log scale.}
    \label{fig:CT_lgH2}
\end{figure}

\textbf{Brush Height and End-monomer Distribution.}
The dependence of brush height on salt concentration under symmetric ion size reduction is shown in \textbf{Figure~\ref{fig:CT_lgH2}}a. 
Across all conditions, the brush height decreases monotonically with increasing salt concentration and gradually levels off in the high-salt regime. 
At low salt concentrations, simultaneous reduction of ion sizes promotes brush contraction, resulting in a lower brush height. 
Notably, in the high-salt regime, the effect becomes non-monotonic: the initially lower brush height for smaller ion sizes is reversed for $c_\mathrm{s} > 0.80$, where smaller ion sizes correspond to a higher brush height.
This crossover indicates the onset of a reentrant swelling regime at very high salt concentrations, which is reflected in the weakened salt-induced collapse and the reduced scaling exponent observed for small ion sizes.
In the log--log representation (\textbf{Figure~\ref{fig:CT_lgH2}}b), the osmotic brush regime shows a scaling exponent of $\alpha \approx 0$. The ion-size dependence of brush height is consistent with that observed for counterion-size variation alone.
In the salted brush regime, decreasing ion size weakens the collapse response to increasing salt concentration, leading to a reduced scaling exponent to $\alpha \approx -0.02$ at $\sigma_{\mathrm{C}}=\sigma_{\mathrm{T}}=0.3$. 
This deviation from classical scaling behavior is associated with enhanced co-ion participation at small ion sizes, which modifies the local ionic environment and partially suppresses brush collapse in the high-salt regime.

We further analyze the distribution of chain-end particles under simultaneous reduction of both ion sizes. The peak position $z_{\max}$ (\textbf{Figure~S10}a) follows the same trend as the brush height $H$, indicating a gradual migration of chain ends toward the grafting surface during brush collapse. 
In the low-salt regime, decreasing ion size shifts $z_{\max}$ toward smaller $z$ values, whereas this trend becomes weaker at higher salt concentrations.
As shown in \textbf{Figure~S10}b and \textbf{S10}c, both $\rho_{\max}$ and FWHM exhibit a transition near $c_\mathrm{s}=0.025$. 
In the low-salt regime, decreasing ion size leads to a reduction in $\rho_{\max}$ and an increase in the FWHM, reflecting a broader chain-end distribution.
At higher salt concentrations, the variations in $\rho_{\max}$ and FWHM with ion size become less pronounced, and the distribution curves gradually converge. 
This suggests that differences in the chain-end distribution shape are progressively weakened, while changes in brush height are primarily associated with shifts in peak position $z_{\max}$.

\begin{figure}[htbp]
   \centering
    \includegraphics[width=1.0\textwidth]{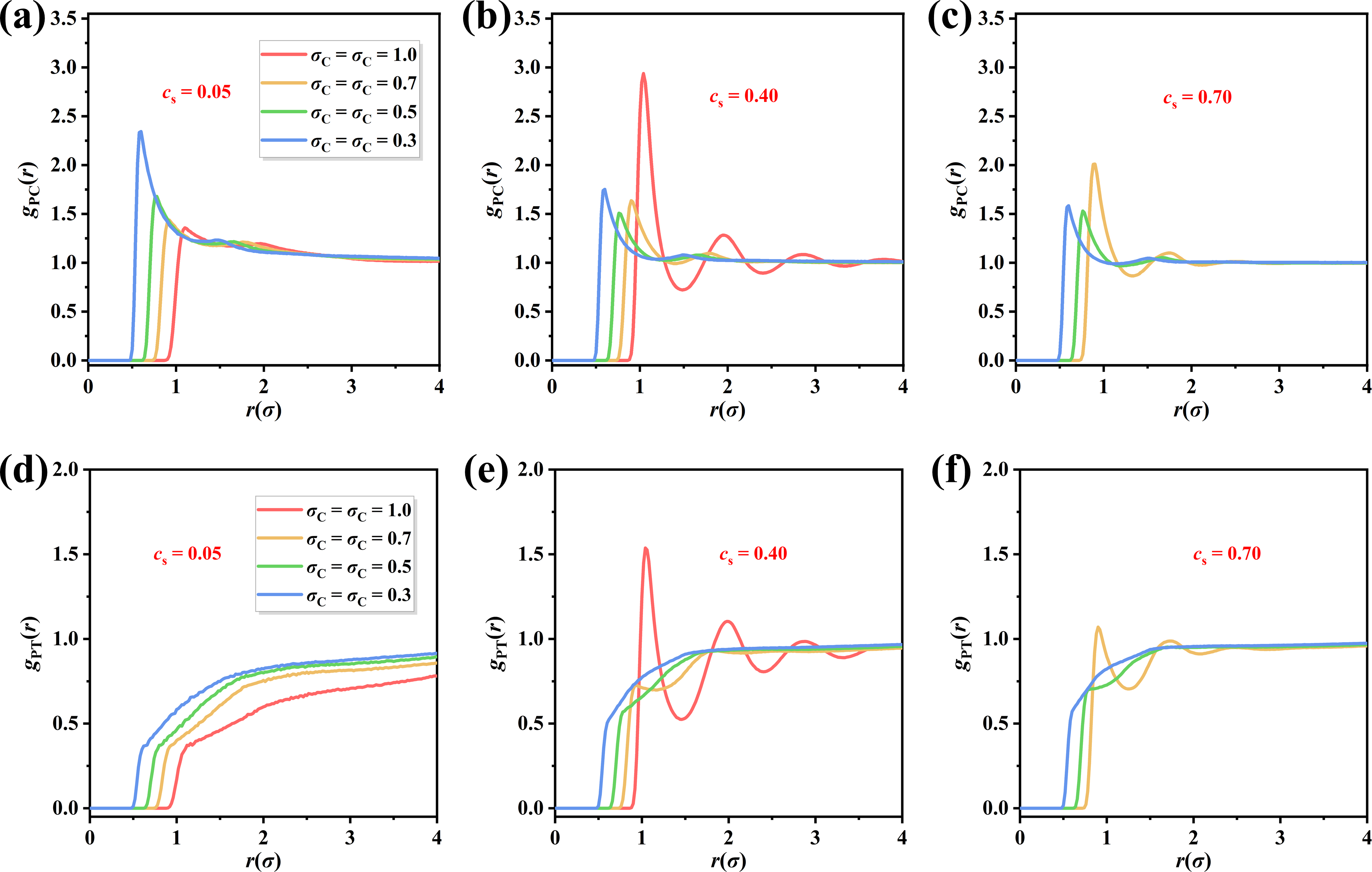}
    \caption{RDFs between PE monomers and counterions (a--c) or co-ions (d--f) when counterions and co-ions decrease simultaneously ($\sigma_\mathrm{C} = \sigma_\mathrm{T}$). The red curve for $c_\mathrm{s}=0.70$ is absent because the maximum accessible salt concentration in the reference system is $c_\mathrm{s}=0.45$.}
   \label{fig:CT_rdf}
\end{figure}

\textbf{Pair Correlation Analysis between PE Monomers and Counterions/Co-ions.}
To clarify how simultaneous reduction of both ion sizes modulates PE chain conformations, we analyze the radial distribution functions between ions and PE monomers.
The P--C interaction is shown in \textbf{Figure~\ref{fig:CT_rdf}}. 
Similar to the case of reducing counterion size alone, a pronounced first-neighbor peak is observed under all conditions, indicating that counterions dominate the local coordination around PE monomers. 
At low salt concentration ($c_\mathrm{s}=0.05$, \textbf{Figure~\ref{fig:CT_rdf}}a), decreasing ion size shifts the first peak toward smaller $r$ and increases its intensity, reflecting enhanced local P--C coordination. 
With increasing salt concentration (\textbf{Figure~\ref{fig:CT_rdf}}b, c and \textbf{Figure~S11a}), the response becomes increasingly size-dependent. 
For larger ion sizes ($\sigma_\mathrm{C}=\sigma_\mathrm{T}=1.0$ and $0.70$), coordination peaks are enhanced, whereas for smaller ion sizes the first peak becomes weaker at high salt concentration. 
The bridging behavior (\textbf{Figure~S12}) exhibits a pronounced dependence on ion size. 
Smaller counterions are less effectively trapped within the brush, leading to reduced intra- and interchain bridging.
Consequently, counterion-mediated correlations between monomers weaken with decreasing ion size, particularly at low salt concentration.
At high salt concentration, this suppression is more pronounced than in the case of reducing counterion size alone.
These behaviors indicate that simultaneous ion-size reduction alters the salt-response of P--C coordination, so that increasing salt concentration does not lead to a monotonic strengthening of local ion–polymer correlations.

Building on the counterion-dominated local coordination structure, we examine the P--T interaction to assess the role of co-ions in charge regulation within the brush.
At low salt concentration ($c_\mathrm{s}=0.05$, \textbf{Figure~\ref{fig:CT_rdf}}d), reducing ion size strengthens the P--T correlation, suggesting enhanced participation of co-ions in local charge organization.
With increasing salt concentration (\textbf{Figure~\ref{fig:CT_rdf}}e, f and \textbf{Figure~S11b}), the P--T interaction becomes increasingly sensitive to ion size. For larger ions, clear coordination peaks develop and grow with salt concentration, indicating stronger co-ion accumulation near PE monomers. 
In contrast, for smaller ions, the peaks are largely smeared out, although the overall distribution shifts upward as salt concentration increases, implying a gradual increase in co-ion involvement.
Importantly, the peak positions of the P--T correlations remain consistent with those of P--C under all conditions, indicating that co-ions do not form independent coordination motifs but instead participate within a counterion-dominated electrostatic environment.

In summary, decreasing ion size enhances the ability of both counterions and co-ions to penetrate into the PE brush. 
Although smaller ions are less favorable for P--C coordination, the enrichment of counterions within the brush still strengthens local charge compensation around PE chains.
In the low-salt regime, these effects preserve a counterion-dominated behavior. 
With increasing salt concentration, co-ions progressively participate in the formation of local electrostatic structures, further weakening P--C interactions.
Meanwhile, the change in counterion coordination suppresses additional brush collapse and drives deviations from single-parameter scaling relations.
At sufficiently high salt concentrations, the combined effects of enhanced ion penetration, altered ion coordination, and modified local charge regulation lead to a partial recovery of brush height, resulting in a reentrant swelling behavior.
Overall, the behavior of PE brushes is governed by the interplay between counterion-dominated coordination and co-ion-mediated charge regulation, which together determine the macroscopic scaling response.

\section{Conclusions}
The response of PE brushes to salt concentration can be described by classical scaling theories, which predict scaling exponents of $\alpha \approx 0$ and $\alpha \approx -1/3$ in the osmotic-brush and salted-brush regimes, respectively. 
However, previous molecular simulation studies have reported deviations from ideal salted-brush scaling, which are generally associated with finite ion size, local electrostatic organization, and non-ideal chain elasticity \cite{kumar2005polyelectrolyte, guptha2014polyelectrolyte}.
Ion-size specificity have been demonstrated to play an important role in determining the microscopic structure of PE brushes \cite{pial2021quantification}. 
In particular, Chen \emph{et al.} developed a cell-model-based theory incorporating inter-monomer electrostatic interactions and proposed a NOEB regime \cite{chen2026scaling}. 
In this regime, the electric double layers surrounding neighboring monomers no longer overlap at high salt concentration, leading to the breakdown of the classical scaling relation and a weaker salt dependence of brush height.
This mechanism provides a consistent interpretation for the reduced scaling exponents widely observed in experiments and simulations.

In this work, we extend the framework within CG molecular simulations by systematically examining ion-size effects in conjunction with salt concentration.
Counterion size primarily governs the scaling response by controlling ion penetration and coordination with PE monomers. 
This effect is most pronounced in the osmotic-brush regime, where the scaling exponent remains close to $\alpha \approx 0$ and the brush height is sensitive to counterion size but weakly dependent on salt concentration. 
At higher salt concentrations, enhanced electrostatic screening reduces the structural differences induced by counterion size, leading to deviations from the asymptotic scaling behavior.
Co-ions, in contrast, influence the system indirectly by modulating ion spatial distributions and the local electrostatic environment. 
Although co-ions do not dominate local coordination, their contribution becomes increasingly non-negligible under realistic ionic conditions. 
When counterion and co-ion sizes are reduced simultaneously, cooperative ion effects emerge: the system is counterion-dominated at low salt concentration, while at high salt concentration, deviations from classical scaling relations arise from co-ion participation, stronger screening, and ion crowding.

Overall, this study provides a microscopic understanding of deviations from classical scaling behavior in PE brushes. 
The present model isolates the steric size effect of monovalent ions within an implicit-solvent CG framework, where ions are represented as charged LJ particles interacting through excluded-volume and Coulombic interactions. 
Consequently, ion size should be interpreted as an effective parameter reflecting both the bare ionic radius and partial hydration effects. 
Ion-specific effects associated with Hofmeister chemistry, including hydration structure, polarizability, and dispersion interactions, are not explicitly considered. 
In addition, other molecular parameters, such as grafting density, chain length, and charge fraction, can also influence the structural response of PE brushes to salt concentration and ion specificity.



\begin{acknowledgement}
This research was supported by the Advanced Materials--National Science and Technology Major Project (2025ZD0614503), the National Natural Science Foundation of China (22373036) and R\&D Program of Guangzhou (2024D03J0007). The computation of this work was supported by Scientific Computing Platform of South China University of Technology.
\end{acknowledgement}

\begin{suppinfo}
Supplemental figures are given in the Supporting Information.
\end{suppinfo}

\bibliography{PEB}

\clearpage
\setcounter{section}{0}
\setcounter{figure}{0}
\renewcommand{\thefigure}{S\arabic{figure}}

\begin{appendix}
\begin{center}
	{\large \textbf{Supplementary Materials}}
\end{center}
\begin{center}
	{\large \textbf{Size Effect of Monovalent Ions on Polyelectrolte Brushes}}
\end{center}

To assess the effect of the simulation box height on the calculated properties, additional simulations were performed with an increased box height of $L_z=100\,\sigma$, while all other simulation parameters were kept identical to those of the reference system.
As shown in \textbf{Figure~S1a}, for the system with $L_z=100\,\sigma$, the PE monomers are mainly located within the region of $z=0$--$25\,\sigma$. 
The corresponding brush height as a function of salt concentration is compared with that for $L_z=60\,\sigma$ in \textbf{Figure~S1b}. Only minor quantitative deviations are observed, while the overall trend remains unchanged. 
These results demonstrate that periodic image interactions along the $z$ direction are negligible in the present simulations and validate the use of $L_z=60\,\sigma$ in this work.

\begin{figure}[htbp]
   \centering
   \includegraphics[width=0.8\textwidth]{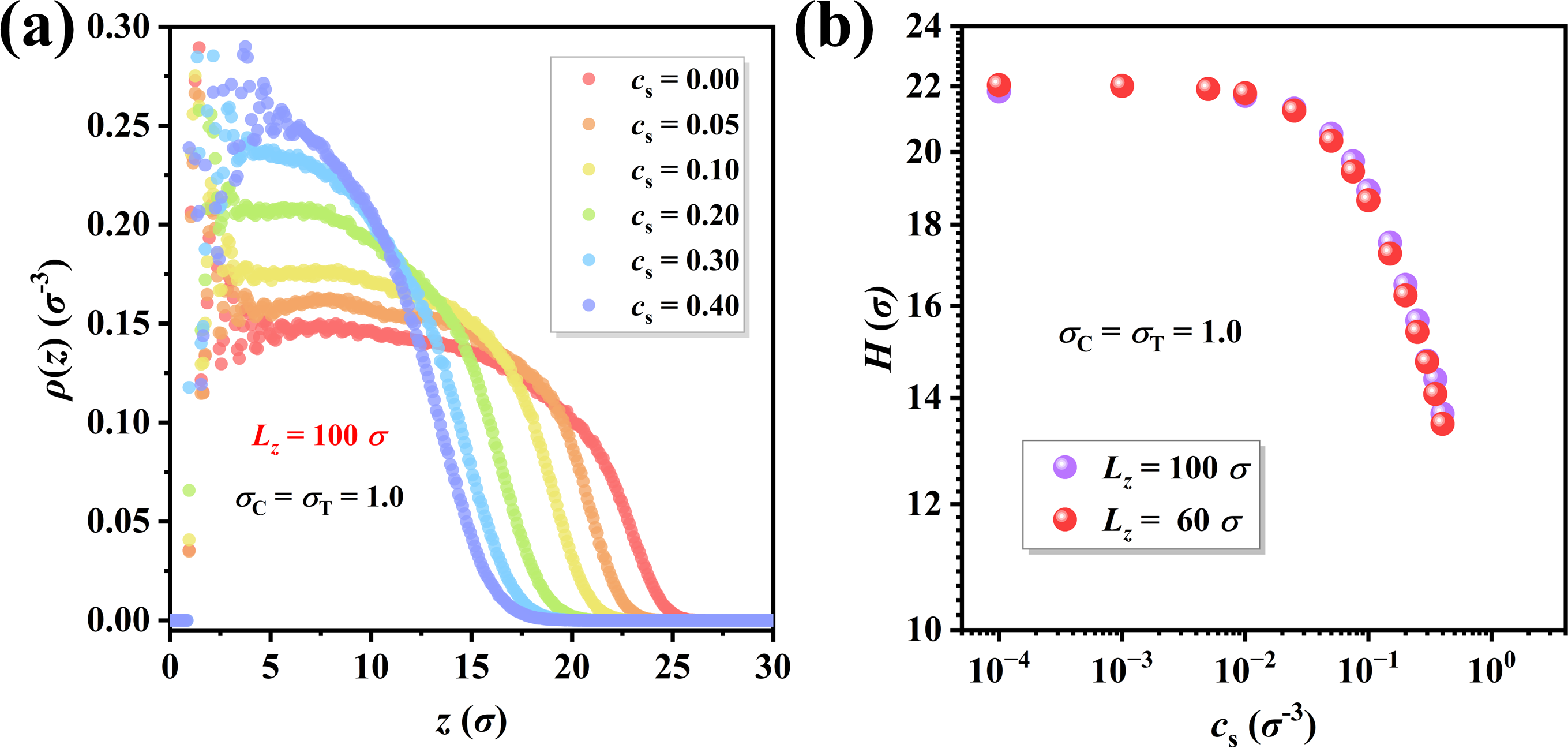}
   \caption{(a) Density profiles of PE monomers along the direction normal to the grafting surface for simulation box heights of $L_z=100\,\sigma$ with $\sigma_\mathrm{C}=\sigma_\mathrm{T}=1.0$. (b) Comparison of the brush height $H$ as a function of salt concentration $c_\mathrm{s}$ for simulation box heights of $L_z=60\,\sigma$ and $L_z=100\,\sigma$ with $\sigma_\mathrm{C}=\sigma_\mathrm{T}=1.0$, plotted on a log--log scale.}
   \label{fig:Box_test}
\end{figure}

\newpage

\begin{table}
   \centering
   \small
   \setlength{\tabcolsep}{20pt}
   \renewcommand{\tablename}{Table}
   \caption{Average brush height $H$ as a function of salt concentration $c_\mathrm{s}$ at $\sigma_\mathrm{T}=1.0$ for different counterion sizes.}
   \label{tab:PEB_C_cs_H}
   \begin{tabular}{ccccc}
      \toprule
      \multirow{2}{*}{\centering $c_\mathrm{s}\,\sigma^3$} & \multicolumn{4}{c}{$H/\sigma$} \\
      \cmidrule(lr){2-5}
      & {$\sigma_\mathrm{C}=1.0$} & {$\sigma_\mathrm{C}=0.7$} & {$\sigma_\mathrm{C}=0.5$} & {$\sigma_\mathrm{C}=0.3$} \\
      \midrule
      0.00\textsuperscript{\emph{a}}   & 22.03 $\pm$ 0.04 & 20.59 $\pm$ 0.00 & 19.50 $\pm$ 0.02 & 18.33 $\pm$ 0.02 \\
      0.001  & 22.01 $\pm$ 0.00 & 20.50 $\pm$ 0.01 & 19.45 $\pm$ 0.01 & 18.26 $\pm$ 0.04 \\
      0.005  & 21.91 $\pm$ 0.03 & 20.41 $\pm$ 0.00 & 19.29 $\pm$ 0.04 & 18.09 $\pm$ 0.01 \\
      0.01   & 21.78 $\pm$ 0.01 & 20.23 $\pm$ 0.03 & 19.05 $\pm$ 0.01 & 17.89 $\pm$ 0.02 \\
      0.025  & 21.24 $\pm$ 0.04 & 19.66 $\pm$ 0.02 & 18.48 $\pm$ 0.02 & 17.24 $\pm$ 0.03 \\
      0.05   & 20.33 $\pm$ 0.04 & 18.80 $\pm$ 0.04 & 17.61 $\pm$ 0.04 & 16.39 $\pm$ 0.04 \\
      0.075  & 19.44 $\pm$ 0.04 & 18.05 $\pm$ 0.04 & 16.97 $\pm$ 0.05 & 15.75 $\pm$ 0.04 \\
      0.10    & 18.64 $\pm$ 0.02 & 17.42 $\pm$ 0.04 & 16.34 $\pm$ 0.03 & 15.24 $\pm$ 0.05 \\
      0.15   & 17.26 $\pm$ 0.08 & 16.33 $\pm$ 0.06 & 15.43 $\pm$ 0.05 & 14.36 $\pm$ 0.04 \\
      0.20    & 16.24 $\pm$ 0.09 & 15.54 $\pm$ 0.05 & 14.62 $\pm$ 0.08 & 13.82 $\pm$ 0.03 \\
      0.25   & 15.41 $\pm$ 0.02 & 14.83 $\pm$ 0.09 & 14.10 $\pm$ 0.04 & 13.25 $\pm$ 0.01 \\
      0.30    & 14.75 $\pm$ 0.05 & 14.24 $\pm$ 0.06 & 13.60 $\pm$ 0.06 & 12.90 $\pm$ 0.04 \\
      0.35   & 14.08 $\pm$ 0.03 & 13.76 $\pm$ 0.07 & 13.26 $\pm$ 0.06 & 12.56 $\pm$ 0.08 \\
      0.40   & 13.48 $\pm$ 0.18 & 13.41 $\pm$ 0.06 & 12.74 $\pm$ 0.06 & 12.28 $\pm$ 0.06 \\
      0.45   & --               & 12.93 $\pm$ 0.07 & 12.47 $\pm$ 0.12 & 12.01 $\pm$ 0.06 \\
      0.50    & --               & 12.55 $\pm$ 0.10 & 12.14 $\pm$ 0.14 & 11.76 $\pm$ 0.19 \\
      0.60    & --               & 12.10 $\pm$ 0.10 & 11.71 $\pm$ 0.15 & 11.32 $\pm$ 0.14 \\
      0.70    & --               & 11.52 $\pm$ 0.06 & 11.04 $\pm$ 0.20 & 11.00 $\pm$ 0.10 \\
      0.80    & --               & 10.97 $\pm$ 0.11 & 10.90 $\pm$ 0.19 & 10.64 $\pm$ 0.15 \\
      \bottomrule
   \end{tabular}
   \begin{flushleft}
   \textsuperscript{\emph{a}} For log-scale representation, the data point at $c_\mathrm{s}=0.00$ is reassigned to $c_\mathrm{s}=10^{-4}$ solely for visualization purposes.
  \end{flushleft}
\end{table}

\newpage

\begin{table}
   \centering
   \small
   \setlength{\tabcolsep}{20pt}
   \renewcommand{\tablename}{Table}
   \caption{Average brush height $H$ as a function of salt concentration $c_\mathrm{s}$ at $\sigma_\mathrm{C}=1.0$ for different co-ion sizes.}
   \label{tab:PEB_T_cs_H}
   \begin{tabular}{ccccc}
      \toprule
      \multirow{2}{*}{\centering $c_\mathrm{s}\,\sigma^3$} & \multicolumn{3}{c}{$H/\sigma$} \\
      \cmidrule(lr){2-4}
      & {$\sigma_\mathrm{T}=0.7$} & {$\sigma_\mathrm{T}=0.5$} & {$\sigma_\mathrm{T}=0.3$} \\
      \midrule
      0.00\textsuperscript{\emph{a}} & 22.05 $\pm$ 0.03 & 22.04 $\pm$ 0.04 & 22.02 $\pm$ 0.06 \\
      0.001  & 22.02 $\pm$ 0.01 & 21.98 $\pm$ 0.03 & 22.02 $\pm$ 0.03 \\
      0.005  & 21.90 $\pm$ 0.00 & 21.92 $\pm$ 0.05 & 21.92 $\pm$ 0.01 \\
      0.01   & 21.79 $\pm$ 0.02 & 21.79 $\pm$ 0.02 & 21.83 $\pm$ 0.03 \\
      0.025  & 21.36 $\pm$ 0.01 & 21.44 $\pm$ 0.01 & 21.49 $\pm$ 0.01 \\
      0.05   & 20.63 $\pm$ 0.07 & 20.81 $\pm$ 0.02 & 21.02 $\pm$ 0.03 \\
      0.075  & 20.02 $\pm$ 0.03 & 20.30 $\pm$ 0.03 & 20.63 $\pm$ 0.02 \\
      0.10   & 19.39 $\pm$ 0.07 & 19.89 $\pm$ 0.02 & 20.22 $\pm$ 0.03 \\
      0.15   & 18.42 $\pm$ 0.03 & 19.11 $\pm$ 0.02 & 19.58 $\pm$ 0.01 \\
      0.20   & 17.60 $\pm$ 0.02 & 18.50 $\pm$ 0.03 & 19.08 $\pm$ 0.05 \\
      0.25   & 17.01 $\pm$ 0.08 & 18.01 $\pm$ 0.03 & 18.61 $\pm$ 0.03 \\
      0.30   & 16.34 $\pm$ 0.06 & 17.54 $\pm$ 0.06 & 18.26 $\pm$ 0.03 \\
      0.35   & 15.80 $\pm$ 0.10 & 17.14 $\pm$ 0.06 & 17.86 $\pm$ 0.06 \\
      0.40   & 15.39 $\pm$ 0.06 & 16.73 $\pm$ 0.05 & 17.54 $\pm$ 0.09 \\
      0.45   & 15.04 $\pm$ 0.13 & 16.54 $\pm$ 0.04 & 17.21 $\pm$ 0.07 \\
      0.50   & 14.71 $\pm$ 0.07 & 16.23 $\pm$ 0.08 & 16.95 $\pm$ 0.17 \\
      0.60   & 14.00 $\pm$ 0.18 & 15.66 $\pm$ 0.16 & 16.36 $\pm$ 0.08 \\
      0.70   & 13.41 $\pm$ 0.24 & 15.28 $\pm$ 0.11 & 15.98 $\pm$ 0.15 \\
      \bottomrule
   \end{tabular}
   \begin{flushleft}
   \textsuperscript{\emph{a}} For log-scale representation, the data point at $c_\mathrm{s}=0.00$ is reassigned to $c_\mathrm{s}=10^{-4}$ solely for visualization purposes.
  \end{flushleft}
\end{table}

\newpage

\begin{table}
   \centering
   \small
   \setlength{\tabcolsep}{20pt}
   \renewcommand{\tablename}{Table}
   \caption{The variation of the average brush height $H$ with salt concentration $c_\mathrm{s}$ under the condition that counterions and co-ions decrease simultaneously ($\sigma_\mathrm{C} = \sigma_\mathrm{T}$).}
   \label{tab:PEB_CT_cs_H}
   \begin{tabular}{ccccc}
      \toprule
      \multirow{2}{*}{\centering $c_\mathrm{s}\,\sigma^3$} & \multicolumn{3}{c}{$H/\sigma$} \\
      \cmidrule(lr){2-4}
      & {$\sigma_\mathrm{C}=\sigma_\mathrm{T}=0.7$} 
      & {$\sigma_\mathrm{C}=\sigma_\mathrm{T}=0.5$} 
      & {$\sigma_\mathrm{C}=\sigma_\mathrm{T}=0.3$} \\
      \midrule
      0.00\textsuperscript{\emph{a}} & 20.58 $\pm$ 0.03 & 19.49 $\pm$ 0.03 & 18.34 $\pm$ 0.04 \\
      0.001  & 20.57 $\pm$ 0.01 & 19.43 $\pm$ 0.03 & 18.28 $\pm$ 0.01 \\
      0.005  & 20.40 $\pm$ 0.03 & 19.32 $\pm$ 0.02 & 18.13 $\pm$ 0.00 \\
      0.01   & 20.25 $\pm$ 0.00 & 19.11 $\pm$ 0.01 & 17.93 $\pm$ 0.01 \\
      0.025  & 19.78 $\pm$ 0.02 & 18.63 $\pm$ 0.04 & 17.48 $\pm$ 0.01 \\
      0.05   & 19.07 $\pm$ 0.03 & 18.05 $\pm$ 0.01 & 16.93 $\pm$ 0.03 \\
      0.075  & 18.50 $\pm$ 0.02 & 17.57 $\pm$ 0.01 & 16.55 $\pm$ 0.01 \\
      0.10   & 17.99 $\pm$ 0.05 & 17.23 $\pm$ 0.04 & 16.20 $\pm$ 0.03 \\
      0.15   & 17.24 $\pm$ 0.03 & 16.66 $\pm$ 0.05 & 15.80 $\pm$ 0.01 \\
      0.20   & 16.58 $\pm$ 0.05 & 16.27 $\pm$ 0.05 & 15.50 $\pm$ 0.01 \\
      0.25   & 16.13 $\pm$ 0.02 & 15.95 $\pm$ 0.01 & 15.29 $\pm$ 0.00 \\
      0.30   & 15.71 $\pm$ 0.05 & 15.73 $\pm$ 0.04 & 15.23 $\pm$ 0.01 \\
      0.35   & 15.40 $\pm$ 0.08 & 15.53 $\pm$ 0.07 & 15.00 $\pm$ 0.01 \\
      0.40   & 15.00 $\pm$ 0.08 & 15.31 $\pm$ 0.06 & 14.93 $\pm$ 0.05 \\
      0.45   & 14.68 $\pm$ 0.06 & 15.24 $\pm$ 0.02 & 14.88 $\pm$ 0.11 \\
      0.50   & 14.56 $\pm$ 0.09 & 15.09 $\pm$ 0.05 & 14.77 $\pm$ 0.05 \\
      0.60   & 14.07 $\pm$ 0.08 & 14.85 $\pm$ 0.04 & 14.66 $\pm$ 0.06 \\
      0.70   & 13.65 $\pm$ 0.11 & 14.69 $\pm$ 0.02 & 14.60 $\pm$ 0.04 \\
      0.80   & 13.32 $\pm$ 0.05 & 14.57 $\pm$ 0.11 & 14.61 $\pm$ 0.03 \\
      0.90   & 13.30 $\pm$ 0.10 & 14.50 $\pm$ 0.04 & 14.48 $\pm$ 0.07 \\
      1.00   & 13.10 $\pm$ 0.23 & 14.42 $\pm$ 0.05 & 14.47 $\pm$ 0.04 \\
      1.10   & 12.86 $\pm$ 0.22 & 14.28 $\pm$ 0.06 & 14.49 $\pm$ 0.04 \\
      1.30   & --               & 14.23 $\pm$ 0.11 & 14.46 $\pm$ 0.07 \\
      1.50   & --               & 14.18 $\pm$ 0.08 & 14.47 $\pm$ 0.05 \\
      1.70   & --               & --               & 14.42 $\pm$ 0.08 \\
      1.90   & --               & --               & 14.27 $\pm$ 0.10 \\
      \bottomrule
   \end{tabular}
   \begin{flushleft}
   \textsuperscript{\emph{a}} For log-scale representation, the data point at $c_\mathrm{s}=0.00$ is reassigned to $c_\mathrm{s}=10^{-4}$ solely for visualization purposes.
  \end{flushleft}
\end{table}

\newpage
\subsection{Decrease $\sigma_{\mathrm{C}}$ and keep $\sigma_{\mathrm{T}} = 1.0$}
\label{sec:PEB_C}

\begin{figure}[htbp]
   \centering
   \includegraphics[width=1.0\textwidth]{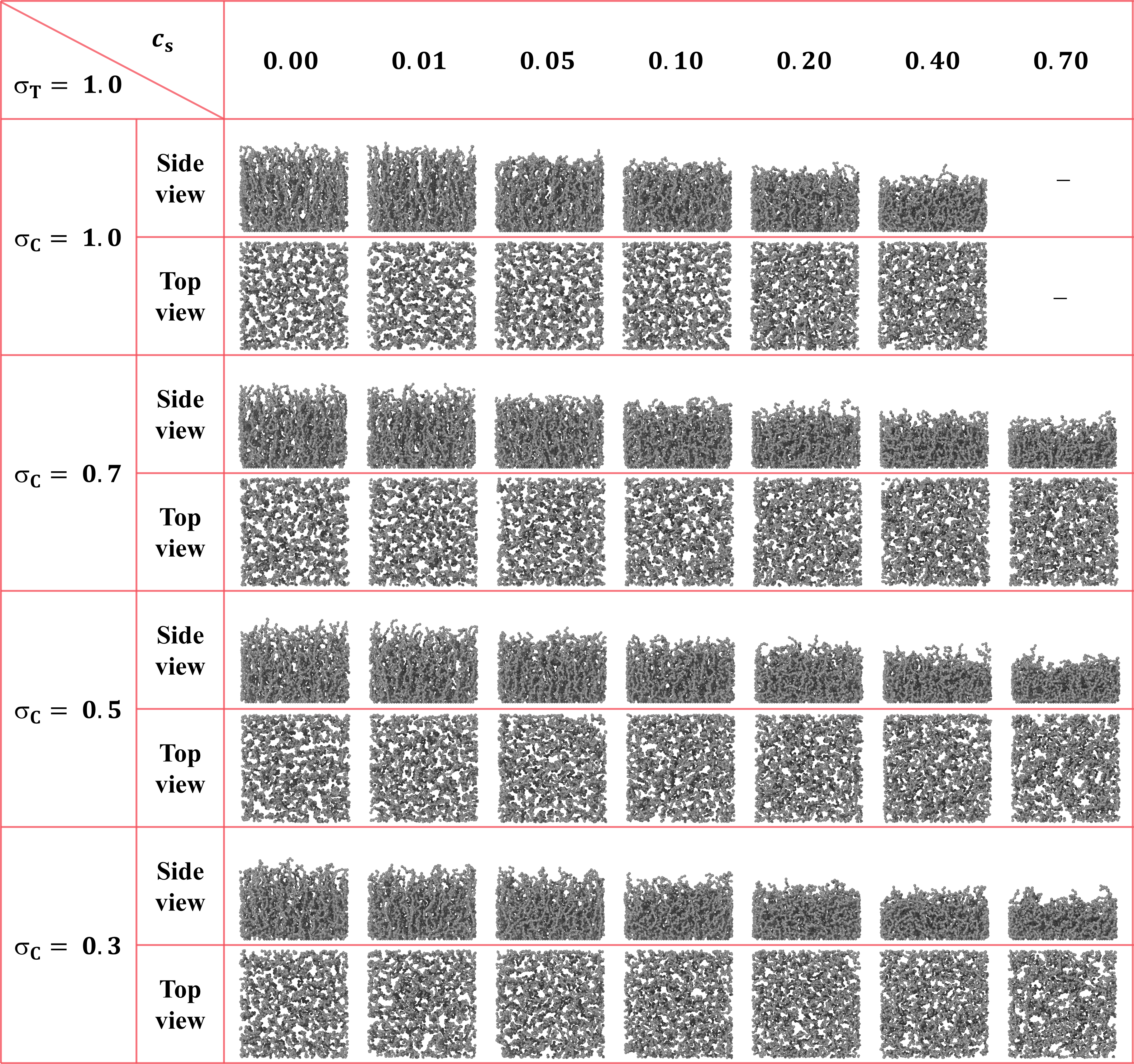}
   \caption{Side and top views illustrating the evolution of PE brush morphologies with salt concentration for different $\sigma_\mathrm{C}$ at $\sigma_\mathrm{T}=1.0$. The views for $c_\mathrm{s}=0.70$ when $\sigma_\mathrm{C}=1.0$ is absent because the maximum accessible salt concentration in the reference system is $c_\mathrm{s}=0.45$.}
   \label{fig:C_snapshot}
\end{figure}

\newpage

\begin{figure}[htbp]
   \centering
   \includegraphics[width=0.45\textwidth]{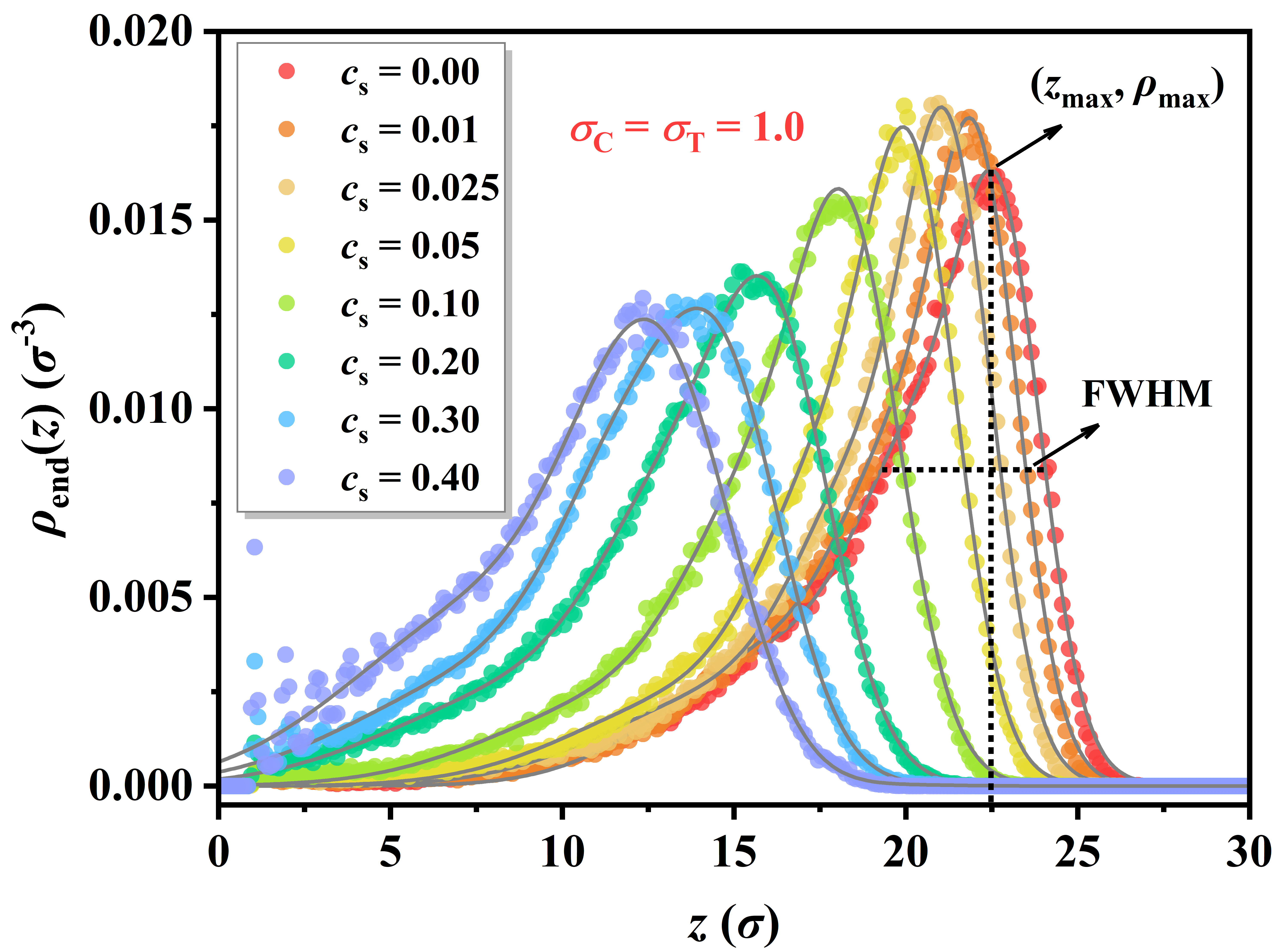}
   \caption{Density profiles of PE chain-end monomers for $\sigma_\mathrm{T}=\sigma_\mathrm{C}=1.0$. The light-gray solid line represents the Gaussian fit used to determine the peak position $z_{\max}$, peak height $\rho_{\max}$, and full width at half maximum (FWHM).}
   \label{fig:end_fit}
\end{figure}

\vspace{2em}

\begin{figure}[htbp]
   \centering
   \includegraphics[width=0.35\textwidth]{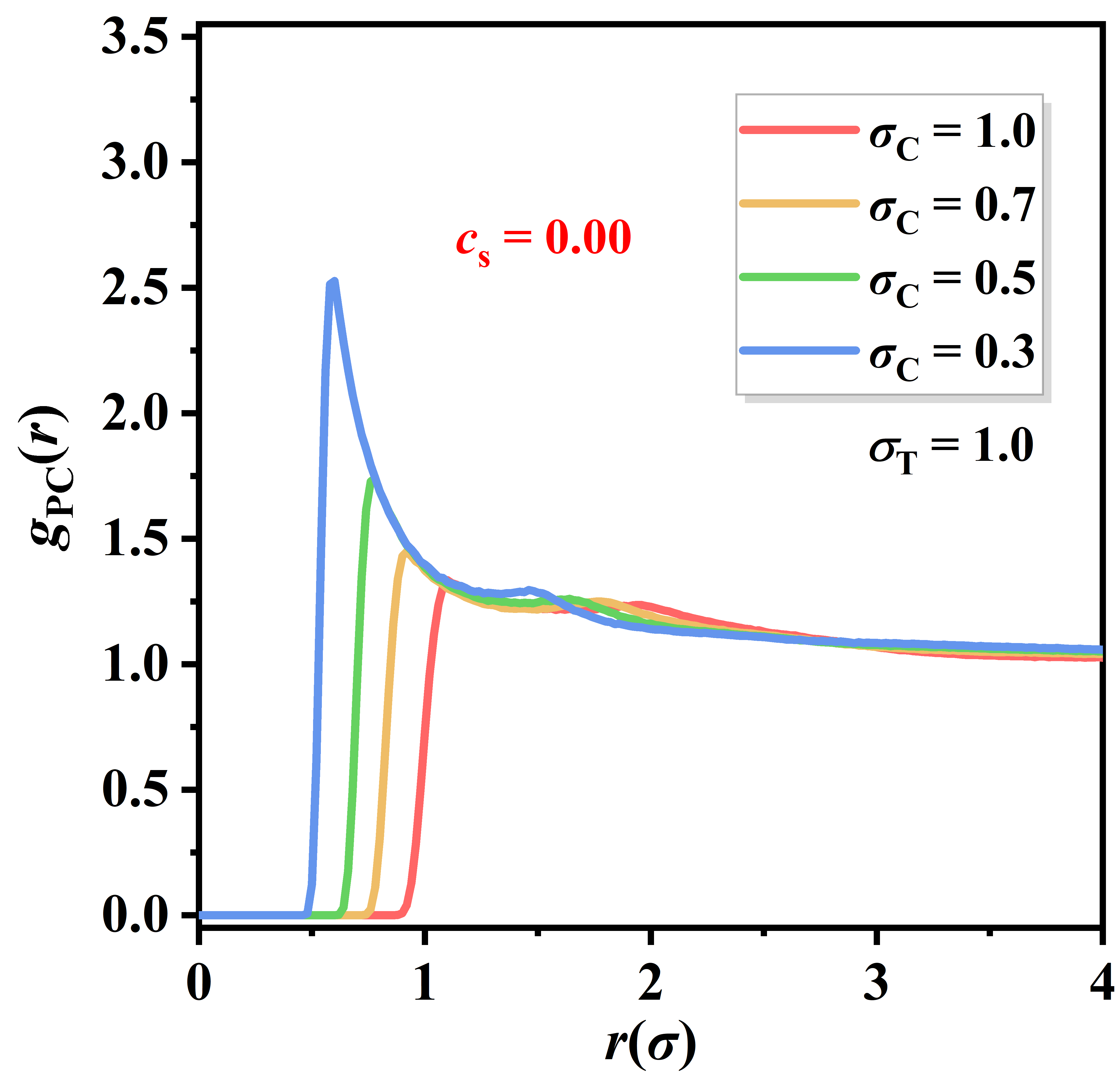}
   \caption{Radial distribution functions (RDFs) between PE monomers and counterions for different $\sigma_\mathrm{C}$ at $\sigma_\mathrm{T}=1.0$ when $c_\mathrm{s}=0.00$.}
   \label{fig:C_rdf_s0}
\end{figure}

\newpage
\subsection{Decrease $\sigma_{\mathrm{T}}$ and keep $\sigma_{\mathrm{C}} = 1.0$}
\label{sec:PEB_T}

\begin{figure}[htbp]
   \centering
   \includegraphics[width=0.4\textwidth]{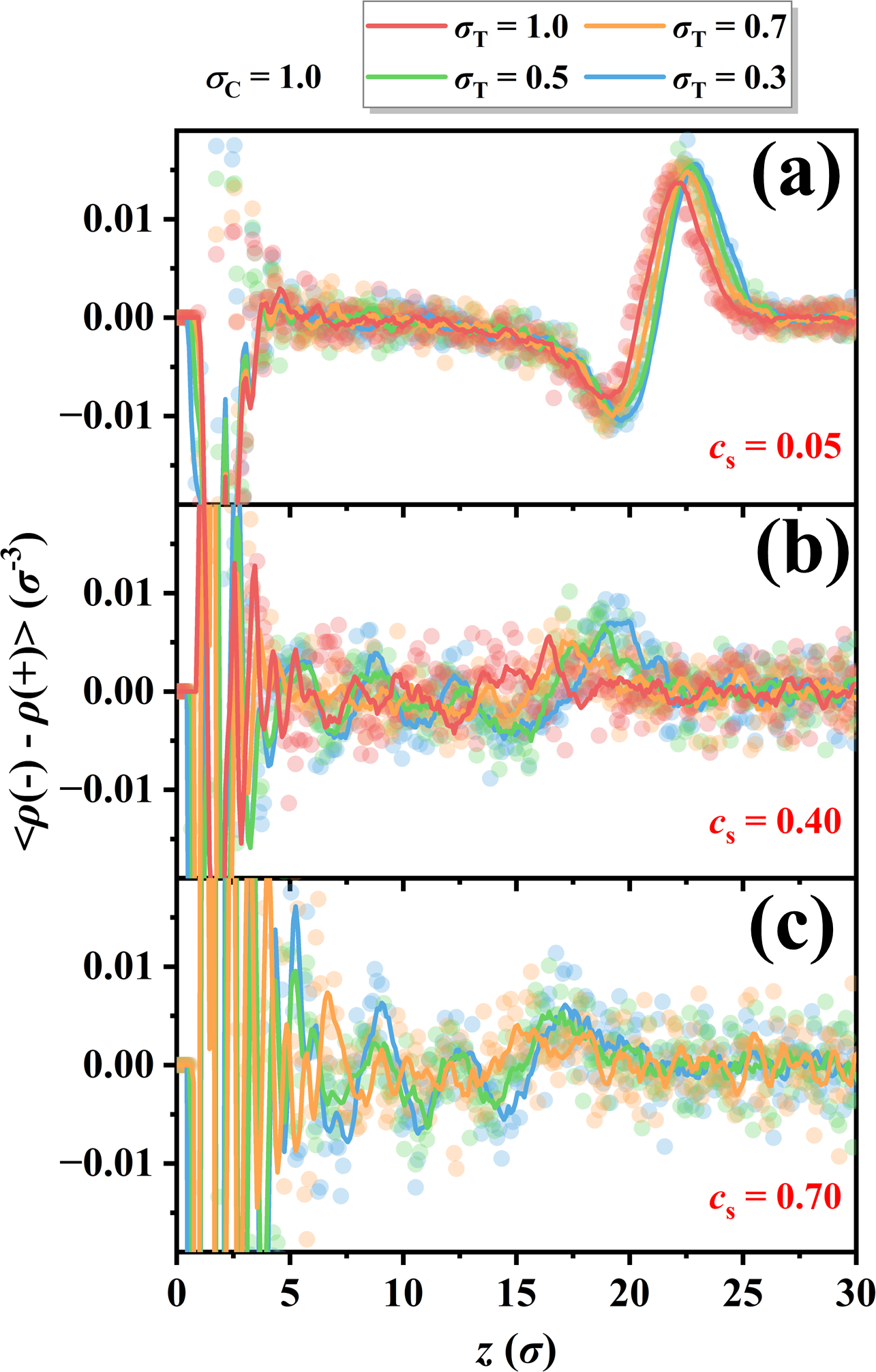}
   \caption{Local net charge density profiles along the direction normal to the grafting surface for different $\sigma_\mathrm{T}$ at $\sigma_\mathrm{C}=1.0$. Panels (a--c) correspond to salt concentrations $c_\mathrm{s}=0.05$, $0.40$, and $0.70$, respectively. Solid lines represent smoothed profiles obtained by convolution, and symbols denote raw data. The red curve for $c_\mathrm{s}=0.70$ is absent because the maximum accessible salt concentration in the reference system is $c_\mathrm{s}=0.45$.}
   \label{fig:T_net_charge}
\end{figure}

\newpage

\begin{figure}[htbp]
   \centering
   \includegraphics[width=1.0\textwidth]{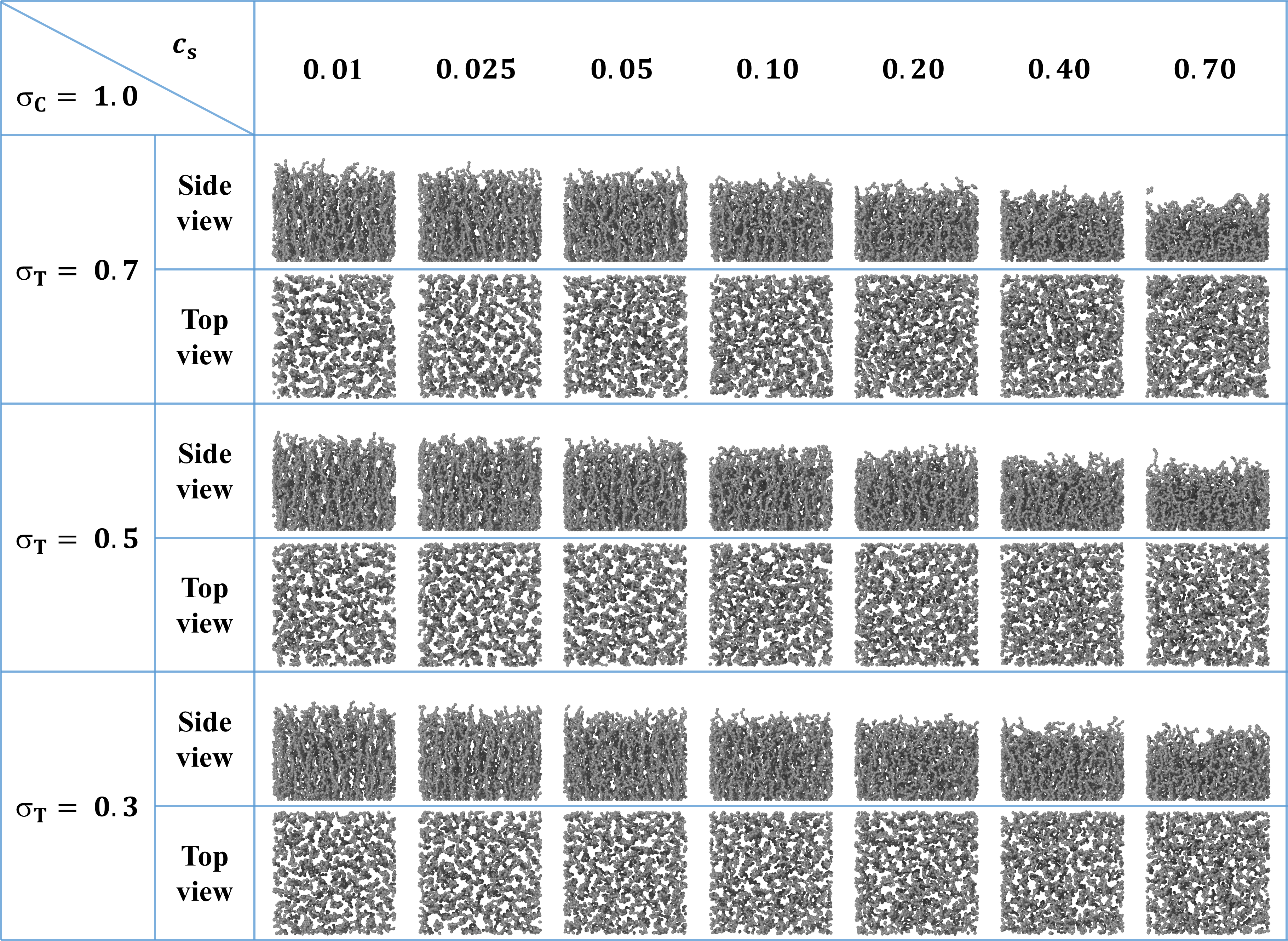}
   \caption{Side and top views illustrating the evolution of PE brush morphologies with salt concentration for different $\sigma_\mathrm{T}$ at $\sigma_\mathrm{C}=1.0$.}
   \label{fig:T_snapshot}
\end{figure}

\newpage

\begin{figure}[htbp]
   \centering
   \includegraphics[width=1.0\textwidth]{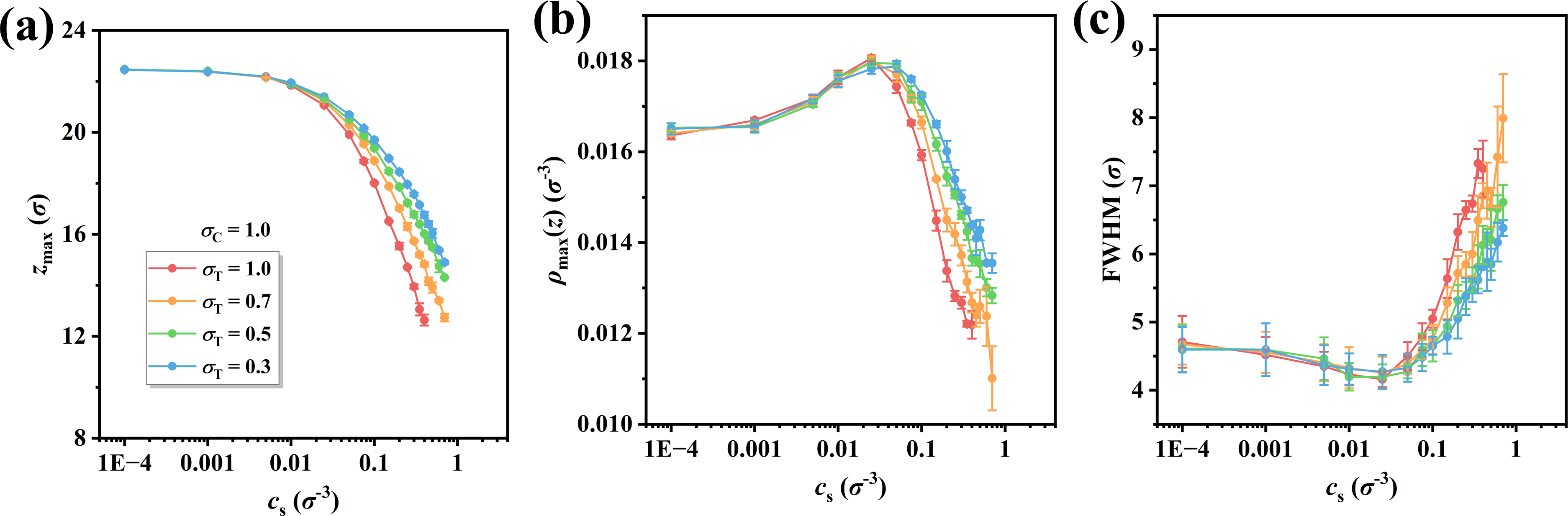}
   \caption{Dependence of characteristic parameters of the PE chain-end monomer density distribution on salt concentration $c_\mathrm{s}$ for different $\sigma_\mathrm{T}$ at $\sigma_\mathrm{C}=1.0$: (a) peak position $z_{\max}$, (b) peak height $\rho_{\max}$, and (c) full width at half maximum (FWHM).}
    \label{fig:T_end_distribution}
\end{figure}

\vspace{2em}

\begin{figure}[htbp]
   \centering
   \includegraphics[width=1.0\textwidth]{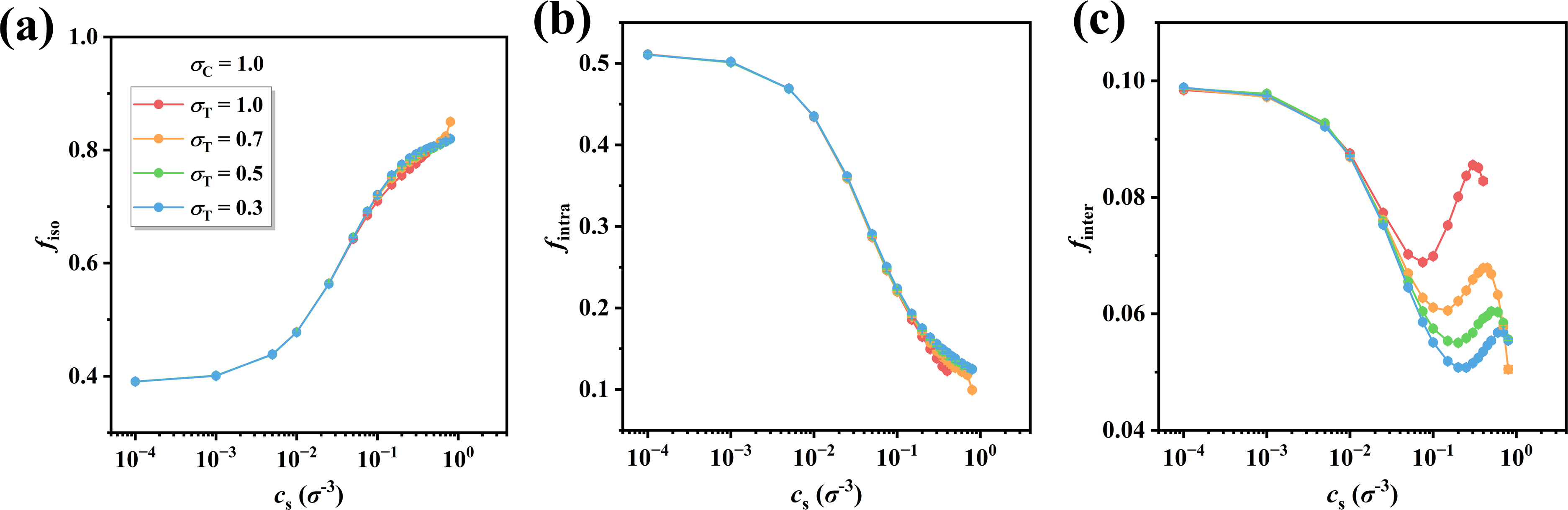}
   \caption{Fractions of different counterion states as functions of salt concentration for different $\sigma_\mathrm{T}$ at $\sigma_\mathrm{C}=1.0$: (a) isolated ($f_{\mathrm{iso}}$), (b) intrachain condensation ($f_{\mathrm{intra}}$), and (c) interchain bridging ($f_{\mathrm{inter}}$). Error bars are smaller than the symbol size.}
   \label{fig:T_counterion}
\end{figure}

\newpage
\subsection{Decrease  $\sigma_\mathrm{C}$ and $\sigma_\mathrm{T}$ simultaneously and keep $\sigma_\mathrm{C} = \sigma_\mathrm{T}$}
\label{sec:PEB_CT}





\begin{figure}[htbp]
   \centering
   \includegraphics[width=1.0\textwidth]{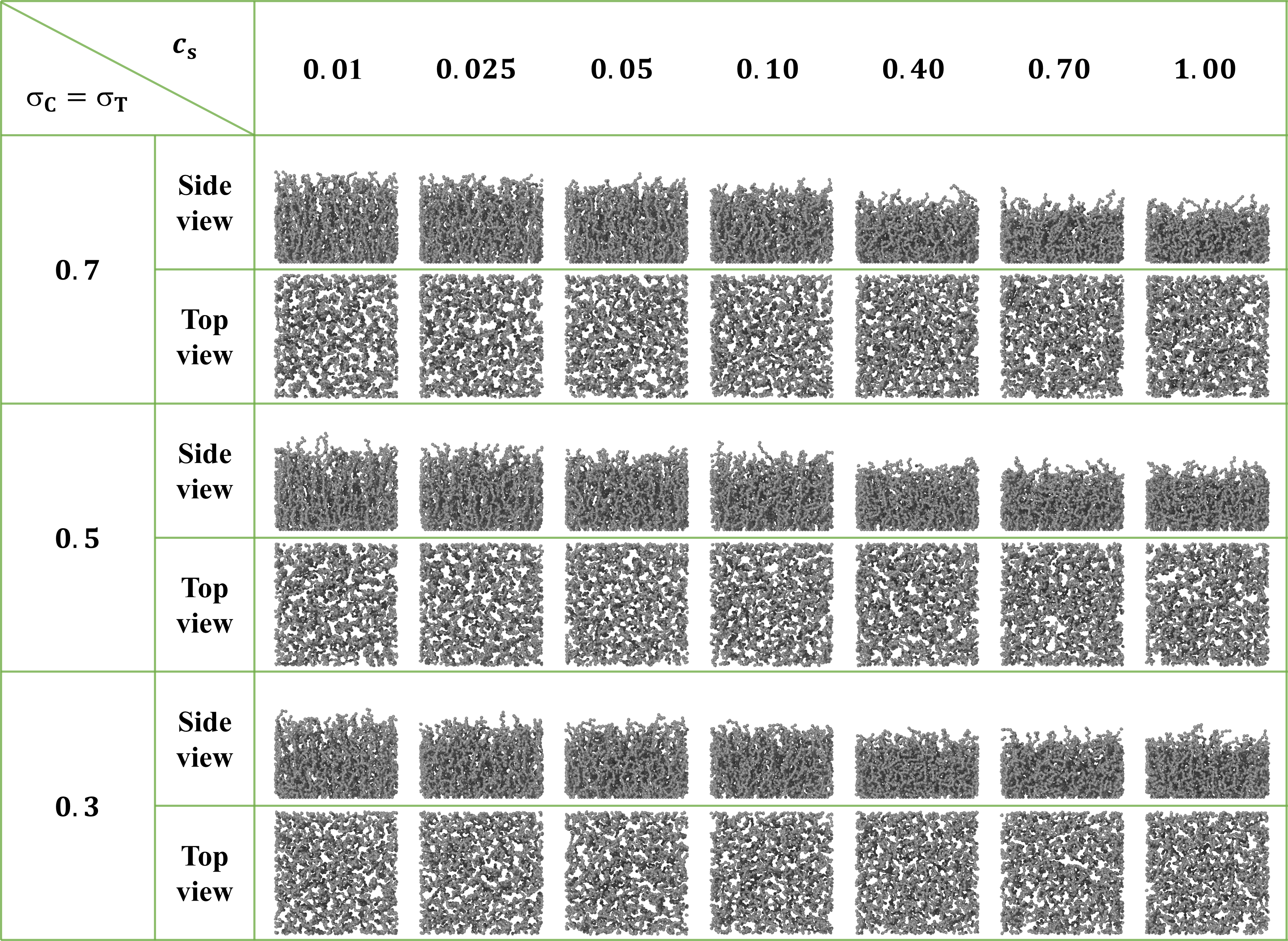}
   \caption{Side and top views illustrating the evolution of PE brush morphologies with salt concentration when counterions and co-ions decrease simultaneously ($\sigma_\mathrm{C} = \sigma_\mathrm{T}$).}
   \label{fig:CT_snapshot}
\end{figure}

\newpage

\begin{figure}[htbp]
   \centering
   \includegraphics[width=1.0\textwidth]{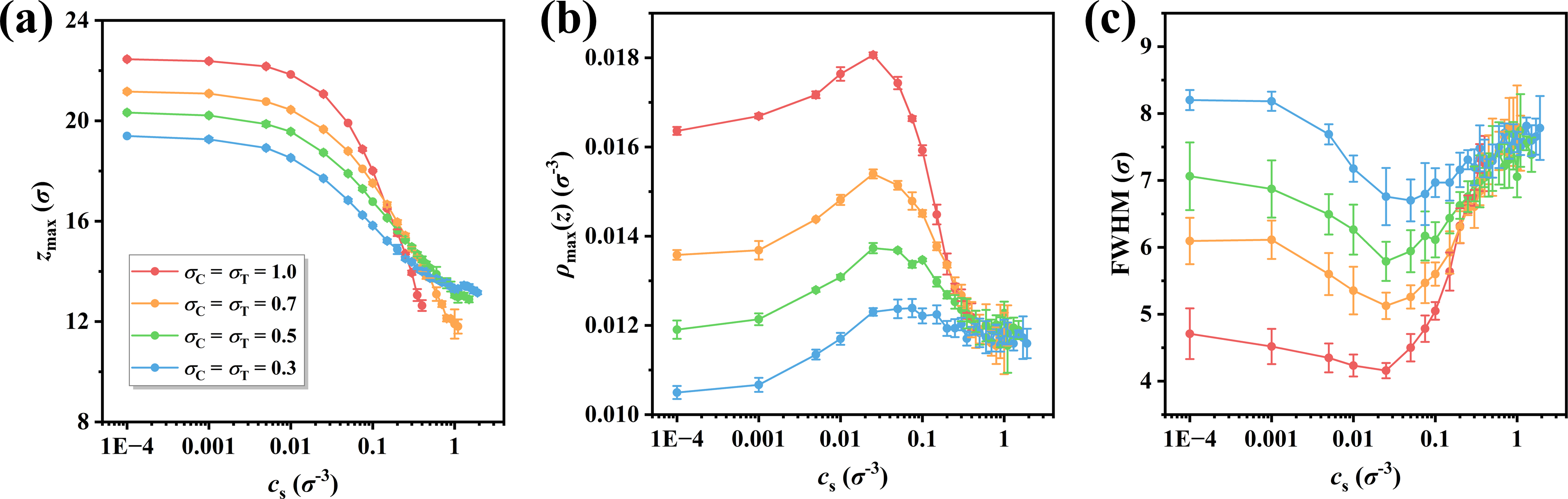}
   \caption{Dependence of characteristic parameters of the PE chain-end monomer density distribution on salt concentration $c_\mathrm{s}$ when counterions and co-ions decrease simultaneously ($\sigma_\mathrm{C} = \sigma_\mathrm{T}$): (a) peak position $z_{\max}$, (b) peak height $\rho_{\max}$, and (c) full width at half maximum (FWHM).}
    \label{fig:CT_end_distribution}
\end{figure}

\vspace{2em}

\begin{figure}[htbp]
   \centering
   \includegraphics[width=0.7\textwidth]{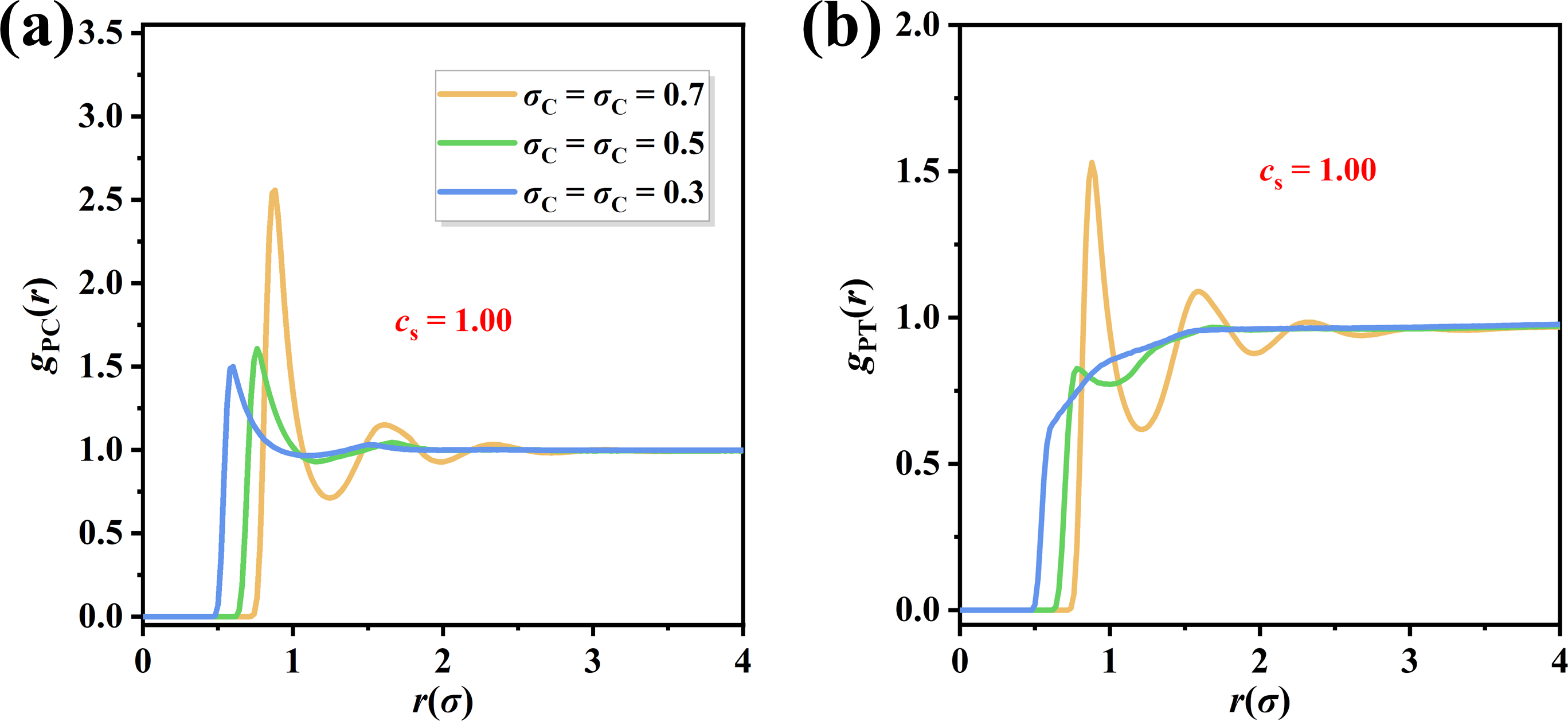}
   \caption{RDFs between PE monomers and counterions (a) or co-ions (b) when counterions and co-ions decrease simultaneously ($\sigma_\mathrm{C} = \sigma_\mathrm{T}$). The red curve for $c_\mathrm{s}=1.00$ is absent because the maximum accessible salt concentration in the reference system is $c_\mathrm{s}=0.45$.}
   \label{fig:CT_rdf_s1.00}
\end{figure}

\newpage

\begin{figure}[htbp]
   \centering
   \includegraphics[width=1.0\textwidth]{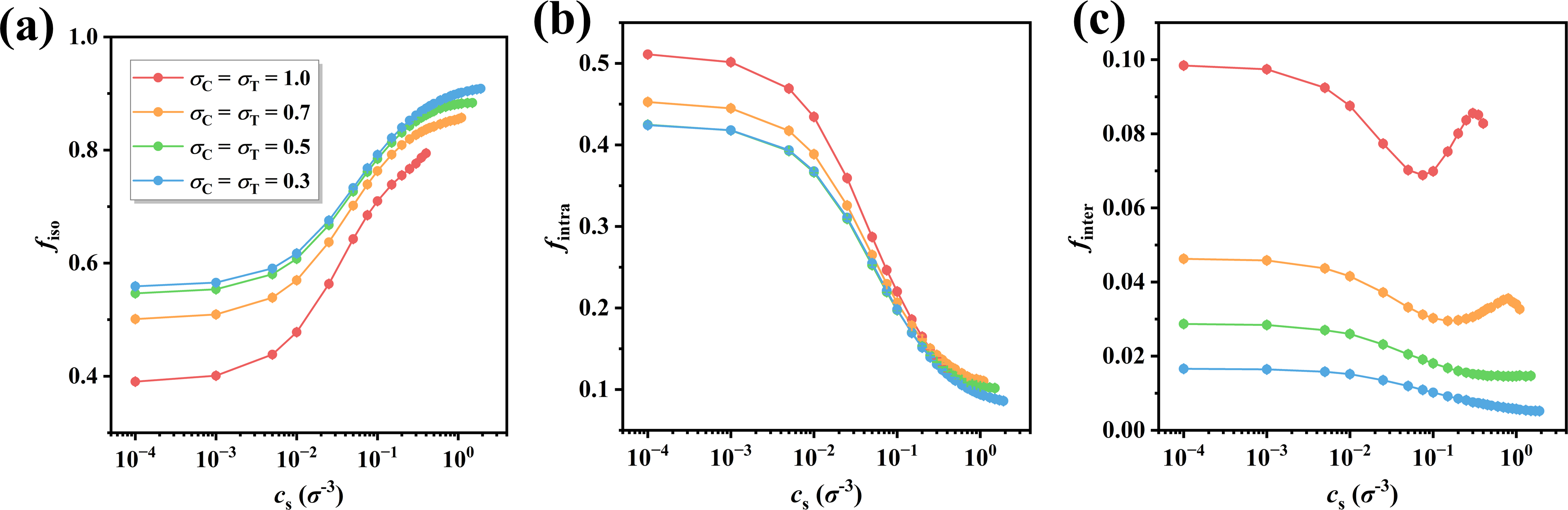}
   \caption{Fractions of different counterion states as functions of salt concentration when counterions and co-ions decrease simultaneously ($\sigma_\mathrm{C} = \sigma_\mathrm{T}$): (a) isolated ($f_{\mathrm{iso}}$), (b) intrachain condensation ($f_{\mathrm{intra}}$), and (c) interchain bridging ($f_{\mathrm{inter}}$). Error bars are smaller than the symbol size.}
   \label{fig:CT_counterion}
\end{figure}

\end{appendix}

\end{document}